\documentclass[10pt,twocolumn,letterpaper]{article}

\usepackage[pagenumbers]{cvpr} 

\usepackage{graphicx}
\usepackage{amsmath}
\usepackage{amssymb}
\usepackage{booktabs}

\usepackage{mathtools}
\usepackage{amsthm}

\usepackage{multirow}
\usepackage{comment}
\usepackage{float} 
\usepackage{siunitx}  
\usepackage{bm}       
\usepackage{subcaption}
\usepackage{caption} %

\usepackage[dvipsnames]{xcolor}

\theoremstyle{plain}

\theoremstyle{definition}

\theoremstyle{remark}

\definecolor{cvprblue}{rgb}{0.21,0.49,0.74}
\usepackage[pagebackref,breaklinks,colorlinks,citecolor=cvprblue]{hyperref}

\usepackage[accsupp]{axessibility}  

\usepackage{xcolor, soul}
\sethlcolor{pink}

\usepackage{pgfplots}
    \usepgfplotslibrary{colorbrewer}
    \pgfplotsset{
        cycle list/Dark2,
        cycle multiindex* list={
            mark list*\nextlist
            Dark2\nextlist
        },
    }
\pgfplotsset{compat=1.14}

\definecolor{ForestGreen}{RGB}{34,139,34}
\definecolor{Cerulean}{rgb}{0.0, 0.48, 0.65}
\definecolor{Cerulean_dark}{rgb}{0.0, 0.30, 0.47}
\definecolor{OrangeRed}{RGB}{245,99,0}
\definecolor{goldenpoppy}{rgb}{0.99, 0.76, 0.0}
\definecolor{goldenpoppy2}{rgb}{0.95, 0.72, 0.0}
\definecolor{skyblue}{rgb}{0.53, 0.81, 0.92}
\definecolor{skyblue2}{rgb}{0.55, 0.83, 0.94}
\definecolor{red-violet}{rgb}{0.78, 0.08, 0.52}
\definecolor{darkcerulean}{rgb}{0.03, 0.33, 0.55}
\definecolor{flamingopink}{rgb}{0.99, 0.59, 0.70}
\definecolor{flamingopink_dark}{rgb}{0.75, 0.32, 0.43}
\definecolor{caribbeangreen}{rgb}{0.0, 0.8, 0.6}
\definecolor{caribbeangreen2}{rgb}{0.0, 0.825, 0.625}
\definecolor{darkpastelpurple}{rgb}{0.69, 0.34, 0.84}
\definecolor{darkpastelpurple2}{rgb}{0.68, 0.33, 0.83}
\definecolor{smokyblack}{rgb}{0.1, 0.1, 0.1}
\definecolor{smokyblack2}{rgb}{0.25, 0.25, 0.25}

\usepackage{standalone}

\usepackage{rotating}

\title{Scanner-Induced Domain Shifts Undermine the\\ Robustness of Pathology Foundation Models}

\author{
Erik Thiringer$^{1}$ \quad
Fredrik K.~Gustafsson$^{1,2}$ \quad
Kajsa~Ledesma~Eriksson$^{1}$ \quad
Mattias~Rantalainen$^{1}$\\[0.5em]
$^{1}$Department of Medical Epidemiology and Biostatistics, Karolinska Institutet, Stockholm, Sweden\\
$^{2}$Department of Engineering Science, University of Oxford, Oxford, United Kingdom\\[0.25em]
{\tt\footnotesize \{erik.thiringer,kajsa.ledesma.eriksson,mattias.rantalainen\}@ki.se, fredrik.gustafsson@eng.ox.ac.uk}
}

\begin{document}

\maketitle


\begin{abstract}
    Pathology foundation models (PFMs) have become a central building block for computational pathology, aiming to provide a general encoder enabling feature extraction from whole-slide images (WSIs) for a wide range of downstream prediction tasks. Despite strong reported performance in benchmark studies, the robustness of PFMs to technical domain shifts commonly encountered in real-world clinical deployment and across studies remains poorly understood. In particular, variability introduced by differences in whole-slide scanner devices represents a common source of variability that has not been characterised as a primary source of domain shift.

In this study, we systematically evaluated the robustness of 14 PFMs to scanner-induced variability. The evaluated models include state-of-the-art PFMs, earlier pathology-specific models trained with self-supervised learning, and a ResNet baseline model trained on natural images. Using a controlled multiscanner dataset comprising 384 breast cancer WSIs scanned on five different whole-slide scanners, we isolate scanner effects independently of biological and laboratory confounders. Robustness is assessed through complementary unsupervised analyses of the embedding space and a set of clinicopathological supervised prediction tasks, including histological grade and routine biomarker predictions from hematoxylin and eosin images.

Our results demonstrate that current state-of-the-art PFMs are not invariant to scanner-induced domain shifts. Most models encode pronounced scanner-specific variability in their embedding spaces, leading to substantial distortions in both global and local feature space across scanners. Although prediction performance largely remains similar when measured by AUC, this apparent robustness masks a critical failure mode: scanner variability systematically alters the embedding space and impacts calibration of downstream model predictions, resulting in scanner-dependent bias that can impact  reliability in, for example, clinical use cases. We further show that robustness to scanner variability is not a simple function of training data scale, model size, or model recency. None of the models provided reliable robustness against scanner-induced variability. The models trained on the most diverse data, here represented by vision-language models, appear to have an  advantage with respect to robustness, while these models did not perform among the top models in the performance evaluation on supervised tasks.   

We conclude that development and evaluation of PFMs requires moving beyond accuracy-centric benchmarks toward explicit evaluation and optimisation of embedding stability and calibration under realistic acquisition variability.

\end{abstract}

\section{Introduction}

\begin{figure*}[t]
    \centering
    \includegraphics[width=0.99\textwidth]{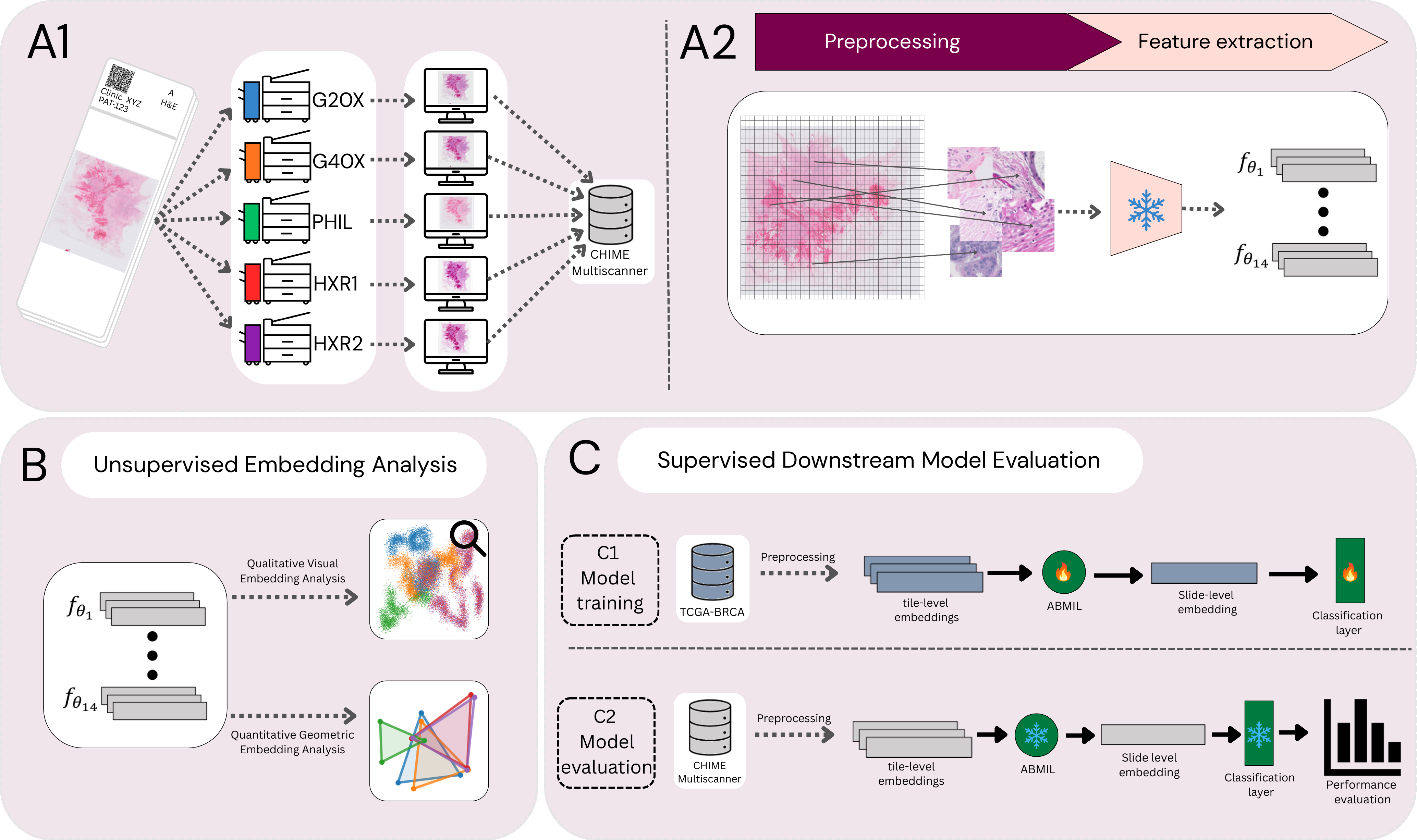}
    \caption{Overview of the study design for evaluating scanner-induced domain shifts in pathology foundation models (PFMs). A1) A total of 384 breast cancer whole-slide images (WSIs) from as many patients were scanned on five different whole-slide scanner devices, forming the CHIME Multiscanner dataset. Each physical slide was scanned using all five devices, thereby isolating scanner-induced variability from all other sources. A2) WSIs from the CHIME Multiscanner and TCGA-BRCA datasets were preprocessed using a standardised workflow and encoded using 14 frozen feature extractors $f_{\theta_i}$, comprising 13 PFMs and a ResNet baseline trained on natural images. B) Scanner-variability robustness is evaluated through complementary qualitative and quantitative unsupervised analyses of feature embedding geometry, capturing both global structure and local neighbourhood consistency across scanners. C) Supervised downstream benchmarking is performed to further assess clinical relevance, with models trained on TCGA-BRCA and evaluated on CHIME Multiscanner. This setup enables systematic evaluation of scanner-induced effects on predictive performance, prediction consistency and calibration stability.}
    \label{fig:experiment_overview}
\end{figure*}

Artificial intelligence (AI) is rapidly transforming both biomedical research and clinical practice, with histopathology image analysis representing one of the most prominent application domains~\citep{Acs2020}. This evolution is driven by the emergence of ever more powerful and efficient AI and machine learning techniques, especially in the form of deep learning, which has proven particularly powerful for image analysis and classification tasks, including medical imaging~\citep{Campanella2019Clinical}. In diagnostic pathology, histological assessment has traditionally relied on manual evaluation of tissue sections under a microscope, and remains a cornerstone of cancer diagnosis and disease characterization. The widespread adoption of digital slide scanning has enabled the generation of high-resolution histopathology whole-slide images (WSIs), facilitating the transition from optical microscopy to digital pathology workflows~\citep{Ghaznavi2013, Song2023}. This transition has, in turn, enabled the systematic application of AI and machine learning techniques to histopathology image data, a field now broadly referred to as computational pathology. 

WSIs contain rich morphological and contextual information that can be leveraged by deep learning models to support a wide range of clinical and research applications~\citep{Arslan2024}, including routine diagnostic assistance, patient stratification, outcome prediction, and prediction of molecular markers directly from hematoxylin and eosin (H\&E)-stained WSIs~\citep{Echle2021, Cifci2022, Wang2022}. For several years, computational pathology pipelines were dominated by convolutional neural networks (CNNs) trained in (weakly) supervised settings~\citep{Srinidhi2021Survey}. More recently, however, vision transformer (ViT)-based architectures trained using self-supervised learning (SSL) on large-scale and diverse datasets have emerged as a new paradigm. These SSL-trained models, commonly referred to as pathology foundation models (PFMs), are intended to serve as general-purpose feature extractors that can be reused across multiple downstream tasks and tissue types.

The adoption of PFMs has fundamentally altered how computational pathology models are constructed. Rather than training task-specific models from scratch, contemporary pipelines typically rely on a pretrained and frozen feature extractor (the PFM), combined with a comparatively lightweight attention-based aggregation model trained for the specific prediction task at hand. This paradigm reduces the need for large amounts of task-specific labeled data and has led to rapid progress across a broad range of applications~\citep{li2025surveycomputationalpathologyfoundation, bilal2025foundationmodelscomputationalpathology}. As a result, numerous PFMs have recently been published and made available to the research community~\citep{conch2024, conchv1.5, Wang2023ctranspath, hoptimus0, hoptimus1, filiot2025_H0-mini, filiot2024phikonv2largepublicfeature, gigapath2024, he2016deep, wang2023retccl, chen2024uni, UNI2h2024, virchow2024, virchow2_2024}, many of which achieve quite similar performance in comparative benchmarking studies~\citep{campanella2024benchmark, neidlinger2024benchmarking, breen2024comprehensive, lee2024benchmarkingpathologyfoundationmodels}. Consequently, distinguishing between released models based solely on standard supervised performance metrics in relatively small sets of tests are becoming increasingly difficult, prompting a shift in focus towards other properties that are critical for clinical deployment, including computational efficiency, generalisability, and robustness.

In particular, a major challenge for clinical deployment is the substantial systemic variability present in real-world histopathology image data, arising from two primary sources. The first source relates to laboratory-specific differences in tissue processing, including fixation protocols, section thickness, staining procedures, and reagent batch effects~\citep{Howard2021SiteBias, Shah2025TissueThickness, Ciompi2017StainNormalization, Dehkharghanian2023BiasedAI}. The second source stems from the image acquisition devices themselves: whole-slide scanners differ across vendors and models in terms of both hardware and software characteristics, leading to systematic differences in colour response, resolution, compression, and image post-processing~\citep{Tellez2019StainNorm, Sikaroudi2022HospitalAgnostic}. Additional slide preparation artifacts such as folds, bubbles and debris may also be present and further contribute to variability~\citep{Taqi2018ArtifactsReview}.

These sources of systemic variability can introduce substantial domain shifts in histopathology image data~\citep{Jahanifar2025DomainGen}. If models are not robust to such domain shifts, their performance might thus degrade when applied to data from unseen laboratories or scanner devices. In clinical settings, this can manifest not only as reduced discriminative accuracy but also as biased predictions or systematic miscalibration, with potentially serious consequences for patient safety. Despite the growing reliance on PFMs as foundational components of computational pathology pipelines, there remains a notable lack of studies evaluating their generalisability and robustness. In particular, the sensitivity of model performance to shifts in scanner devices, staining protocols and other preprocessing factors is still poorly understood.

Importantly, robustness in this context extends beyond maintaining high ranking-based performance metrics such as area under the receiver operating characteristic curve (AUC). Even when discriminative performance appears stable, scanner-induced shifts may alter the space of learned feature representations and introduce bias in predicted probability distributions, leading to scanner-dependent decision thresholds. Such effects are largely invisible in standard benchmarking protocols and have so far not been evaluated systematically, despite their direct relevance for clinical deployment.

The primary objective of this study is therefore to systematically evaluate the robustness of a diverse set of previously published PFMs with respect to scanner-induced variability. Focusing on whole-slide scanner devices as a major and clinically relevant source of domain shift, we present a controlled multiscanner benchmark that isolates scanner effects independently of biological and laboratory confounders. Using a real-world dataset in which identical tissue specimens are scanned on multiple devices, we assess robustness through a combination of unsupervised embedding space analyses and supervised downstream evaluations. By jointly analysing embedding geometry, prediction consistency and probability calibration, our study aims to uncover failure modes that are not captured by accuracy-centric benchmarks, and to identify architectural and training strategies that promote scanner-robust feature representations. An overview of the study design is shown in Figure~\ref{fig:experiment_overview}, and the main contributions of this work are summarised below.

\begin{itemize}
    \item \textbf{Controlled Multiscanner Evaluation:}  
    We present a controlled evaluation of scanner-induced domain shifts using the in-house CHIME Multiscanner dataset, comprising 384 breast cancer H\&E WSIs scanned on five different whole-slide scanners. By analysing identical tissue specimens across scanner devices from a single clinical site, our study isolates scanner-induced effects from all other sources of variation.
    
    \vspace{1.0mm}
    \item \textbf{Comprehensive Evaluation of Feature Extractors:}
    We evaluate the robustness of 14 previously published and widely used feature extractors to scanner-induced variability, spanning several generations and design paradigms. The evaluated models include current state-of-the-art ViT-based PFMs, vision-language PFMs, earlier pathology-specific models trained with self-supervised learning, and a ResNet baseline trained on natural images.

    \vspace{1.0mm}
    \item \textbf{Quantitative Geometric Embedding Analysis:}
    We systematically characterise scanner sensitivity in the learned feature embedding space using complementary geometric metrics, including pairwise cosine distance, Mantel correlation, and neighbourhood consistency. This analysis reveals that the majority of PFMs encode pronounced scanner-specific signatures that substantially distort both local and global embedding geometry.

    \vspace{1.0mm}
    \item \textbf{Benchmarking on Clinical Downstream Tasks:} 
    We benchmark downstream performance on five breast cancer clinicopathological prediction tasks, training on TCGA-BRCA and evaluating on our multiscanner dataset. This enables direct assessment of how scanner-induced representation shifts propagate to discriminative performance, prediction consistency, and probability calibration under realistic acquisition variability.

    \vspace{1.0mm}
    \item \textbf{Identification of Hidden Failure Modes:}  
    We show that, although discriminative performance often remains strong when measured by AUC, this apparent robustness masks a critical failure mode: scanner variability systematically alters the embedding space and downstream probability calibration, resulting in scanner-dependent decision thresholds that can compromise reliability in clinical deployment.
    
    \vspace{1.0mm}
    \item \textbf{Insights for Robust PFM Development:}  
    We demonstrate that robustness to scanner-induced domain shifts is not a simple function of training data scale, model size, or model recency. None of the models provided reliable robustness against scanner-induced variability. The models trained on the most diverse data, here represented by vision-language models, appear to have an  advantage with respect to robustness, while they did not perform among the top models in the supervised tasks. 
\end{itemize}

\section{Related Work}

Domain shifts, including scanner variability, have been identified as a key challenge for clinical deployment of AI-based histopathology models that use PFMs~\citep{bilal2025foundationmodelscomputationalpathology}. Prior work has shown that deep learning models can be trained to identify the acquisition site of histopathology images, indicating the presence of site-specific characteristics in the image data~\citep{Dehkharghanian2023BiasedAI}. It has also been reported that variations in laboratory protocols (such as tissue section thickness) across labs can alter computed image features, underscoring how site-specific practices can introduce bias in model outputs~\citep{Shah2025TissueThickness}. Moreover, \citet{komen2024batch} demonstrated that larger PFMs capture distinct site-related signatures that can lead to bias in predictions, with similar findings reported also by \citet{Dehkharghanian2023BiasedAI} and \citet{lin2025institutionbias}. A certain type of robustness has however also been observed, as \citet{Elphick2024RotationInvariance} reported rotational invariance in the embeddings produced by some foundation models.

Comparative benchmarking studies of PFMs have been published recently~\citep{campanella2024benchmark, neidlinger2024benchmarking, breen2024comprehensive}, but none have directly compared performance on the same set of WSIs scanned using different whole-slide scanners. As illustrated in Figure~\ref{fig:mosaic}, scanner-induced colour variation can be substantial in practice. Understanding the impact of such scanner-related variability on PFMs represents a critical knowledge gap with significant implications for clinical deployment as well as for multi-institutional research studies.

Recent concurrent work has begun to explore scanner-induced variability more explicitly. \citet{ryu2025scorpion} introduced a multiscanner dataset digitised using five different scanners, but their dataset consists of small image \emph{patches} extracted from only 48 WSIs, and does not include any evaluation of pretrained PFMs. Instead, their study investigates the effectiveness of style-based augmentation and consistency-based losses when training a tissue segmentation model from scratch.

\citet{carloni2025pathologyfoundationmodelsscanner} evaluated five PFMs, all of which are included in our study, and demonstrated that all tested models exhibit sensitivity to scanner variation.. They further proposed a new loss function to train MIL models, utilizing a curated multiscanner dataset, that improved robustness to scanner variation. However, this approach depends on access to a multiscanner dataset, and because the optimisation is applied only during downstream model training, the underlying tile-level embeddings produced by the PFMs themselves remain inherently non-robust.

Similarly, \citet{Chai2025.08.18.670932} investigated the impact of both tissue-staining and scanner variability for six different PFMs and a ResNet-50 baseline. Their analysis focused on predictive performance, mainly using accuracy, and highlighted that performance degradation caused by domain shifts could be partially mitigated by incorporating a small number of stain-varied slides into the downstream model training.

While these concurrent studies collectively confirm the presence of scanner-induced domain shifts, our work distinguishes itself through a more granular and clinically oriented evaluation framework. In contrast to \citet{carloni2025pathologyfoundationmodelsscanner}, who primarily propose a mitigation strategy that requires curated multiscanner training data, we rigorously assess the intrinsic, out-of-the-box stability of PFM embedding spaces using geometric consistency metrics. Moreover, distinct from ~\citet{Chai2025.08.18.670932}, who primarily evaluate classification accuracy and data efficiency in a rare tumour setting, we isolate scanner-induced effects in a common diagnostic setting (breast cancer) and explicitly examine impact on clinical reliability.

Specifically, by benchmarking 14 diverse feature extractors using our CHIME Multiscanner dataset, we conduct a large-scale, controlled evaluation demonstrating that scanner variability can induce systematic shifts in predicted probabilities even when discriminative performance, as measured by AUC, appears stable. These probability shifts result in miscalibration and compromised clinical decision thresholds, revealing a critical failure mode that is largely overlooked in prior accuracy-centric benchmarking studies.

\section{Methods}
This study utilizes the in-house CHIME Multiscanner dataset together with TCGA-BRCA to evaluate scanner-variability robustness of 14 different feature extractors, including a range of recent PFMs. The datasets are introduced in Section~\ref{subsection:methods_datasets}, our standardised WSI preprocessing workflow is outlined in Section~\ref{subsection:methods_preprocessing}, while the different types of feature extractors are detailed in Section~\ref{subsection:methods_feature-extractors}.

\begin{figure*}[t]
    \centering
    \includegraphics[width=0.99\textwidth]{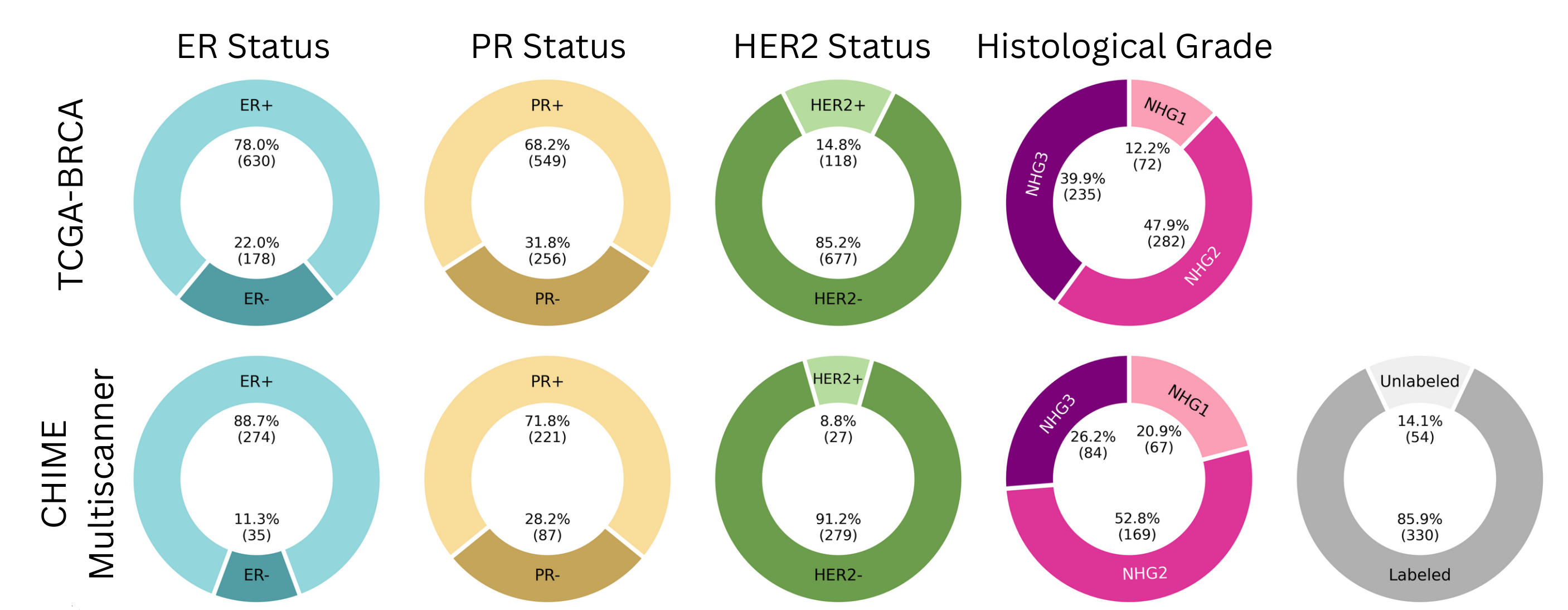}
    \caption{Summary of dataset composition for CHIME Multiscanner and TCGA-BRCA. Donut charts illustrate the class distribution for the clinical biomarkers (ER, PR, HER2) and histological grades (NHG1, NHG2, NHG3) within the training (TCGA-BRCA) and evaluation (CHIME Multiscanner) cohorts. For CHIME Multiscanner, the proportion of labeled and unlabeled WSIs is also shown, as unlabeled data is utilized for the unsupervised embedding analysis.}
    \label{fig:label_proportion}
\end{figure*}

\subsection{Datasets}
\label{subsection:methods_datasets}

\paragraph{CHIME Multiscanner}
This in-house dataset consists of 384 histopathology slides of H\&E-stained breast tumour sections from the same number of patients, from a single medical center (Södersjukhuset in Stockholm, Sweden), that were diagnosed in 2015. The set of slides was scanned on five whole-slide scanner devices from three different manufacturers, see Figure~\ref{fig:experiment_overview}A1 and Table~\ref{scanner}. The total number of digitized WSIs is thus 1920 (384 slides $\times$ 5 scanners).

\begin{figure}[t]
    \centering
    \includegraphics[width=1.0\linewidth]{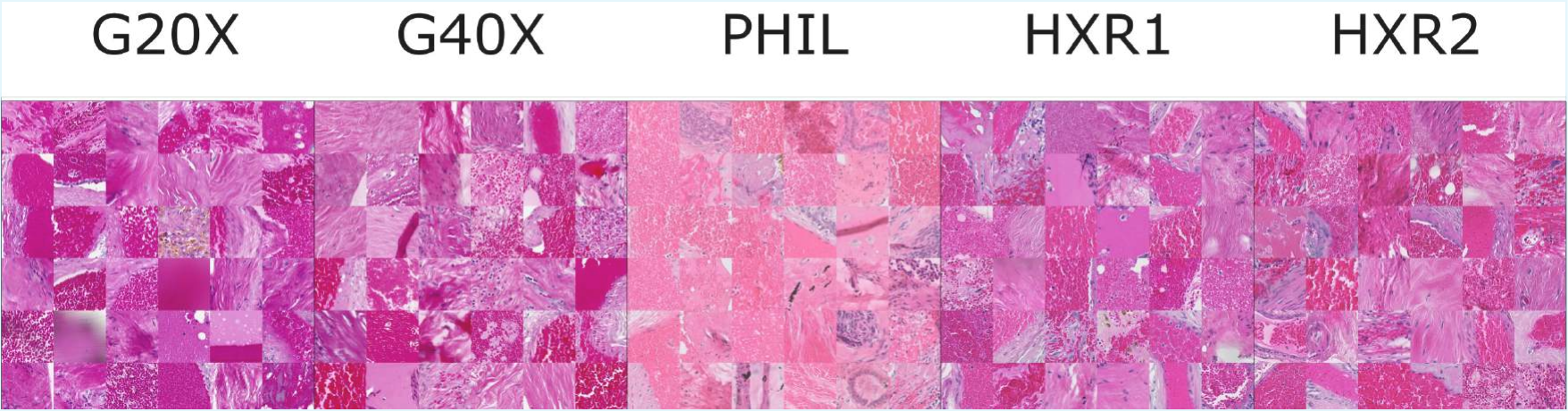}\vspace{-1.0mm}
    \caption{Representative mosaic of tissue tiles sampled from unique WSIs in the CHIME Multiscanner dataset, grouped by scanner device, illustrating scanner-dependent visual variability.}
    \label{fig:mosaic}
\end{figure}

All image acquisition was performed at $40\times$ magnification, with the exception of the Grundium Ocus20 scanner (G20X), which has a maximum magnification of $20\times$. Figure~\ref{fig:mosaic} displays a mosaic of tissue tiles sampled from the multiscanner dataset, illustrating the scanner-induced colour variation. While tiles from HRX1 and HRX2 are perceptually indistinguishable, as expected given that these are two different devices of the same scanner model, there is a significant colour shift e.g. between G40X and PHIL.

The dataset also contains clinicopathological information in the form of Nottingham histological grade (NHG) and routine biomarker status (ER, PR, and HER2) from clinical routine assessments for most patients, see Figure~\ref{fig:label_proportion} for an overview of the distribution of available clinical factors. CHIME Multiscanner is used to characterise scanner-related variability in the context of both unsupervised analysis of the feature extractor embedding space, and in the supervised downstream model evaluation (prediction of biomarkers and histological grade).

\begin{table}[t] 
    \caption{The five whole-slide scanner devices utilized in this study. HXR1 and HXR2 are two different devices of the same model.}\vspace{-2.0mm}
    \label{scanner}
    \centering
	\resizebox{1.0\linewidth}{!}{%
        \begin{tabular}{cllc}
        \toprule
        \textbf{Short Name} &\textbf{Manufacturer} &\textbf{Model Name} &\textbf{Magnification} \\ 
        \midrule
        G20X &Grundium &Ocus20 &$20\times$ \\ 
        G40X &Grundium &Ocus40 &$40\times$ \\
        PHIL &Philips &Ultra Fast Scanner &$40\times$ \\
        HXR1 &Hamamatsu &NanoZoomer-XR &$40\times$ \\
        HXR2 &Hamamatsu &NanoZoomer-XR &$40\times$ \\
        \bottomrule
        \end{tabular}
	}
\end{table}
\paragraph{TCGA-BRCA}
In total, TCGA-BRCA consists of 1$\thinspace$133 diagnostic WSIs from 1$\thinspace$098 unique patients, collected from more than 40 different sites\footnote{\href{https://gdc.cancer.gov/resources-tcga-users/tcga-code-tables/tissue-source-site-codes}{https://gdc.cancer.gov/resources-tcga-users/tcga-code-tables/tissue-source-site-codes}}.
In our study we utilized the subset of TCGA-BRCA from 2012~\citep{TCGA2012BRCA}, available at \href{https://www.cbioportal.org/study/clinicalData?id=brca_tcga_pub}{cBioPortal,} 
which includes both molecular profiling information, histopathology WSIs and clinicopathological data. This subset consists of 825 primary piece breast cancer samples from the same number of patients, and it serves as the training cohort for our supervised downstream model evaluation, as illustrated in Figure~\ref{fig:experiment_overview}C. The clinical factors from CHIME Multiscanner (ER, PR and HER2 status, NHG) are available for a subset of TCGA-BRCA WSIs, see Figure~\ref{fig:label_proportion} for details. 

\begin{table*}[t]
    \centering
    \caption{Overview of the 14 feature extractors evaluated in this study, including a range of recent PFMs and a ResNet baseline. For each model, we report release date, self-supervised learning approach, architecture, embedding dimensionality, training WSI magnification, approximate training set size, and data source. Missing entries (-) indicate information that is not publicly available or not applicable.}\vspace{-2.0mm}
    \renewcommand{\arraystretch}{1.4}
    \resizebox{\textwidth}{!}{%
        \begin{tabular}{lccccccc}
        \toprule
        \textbf{Name} & \textbf{Released} & \textbf{SSL Approach} & \textbf{Architecture} & \textbf{Embedding Size} & \textbf{Magnification} & \textbf{Training Size (WSIs)}  & \textbf{Data Source} \\
        \midrule
        CONCH ~\citep{conch2024} & Jul 2023 & iBOT $\rightarrow$ CoCa & ViT-B/16 & 512 &$20\times$ & $\approx$ 20K $\rightarrow$ 1.2 million pairs &  Public $\rightarrow$ Mixed  \\ 
        CONCHv1.5 ~\citep{conchv1.5} & Oct 2024 & - & ViT-L & 768 & -  & -  & - \\
        CTransPath ~\citep{Wang2023ctranspath} & Jul 2022 & SRCL & CNN + Swin-T  & 768& $20\times$  & $>$ 30K  & Public  \\
        H-Optimus-0 ~\citep{hoptimus0}& Jul 2024 & DINOv2 & ViT-G/14 & 1536 & $20\times$ & 500K  & Private \\
        H-Optimus-1 ~\citep{hoptimus1}& Mar 2025 & - & ViT-G/14 & 1536 & $20\times$  & $>$ 1 million  & Private  \\
        H0-mini ~\citep{filiot2025_H0-mini} & Jan 2025 & DINOv2 & ViT-B & 768 & $20\times$  & 500K + 6k  & Mixed  \\
        Phikon-v2 ~\citep{filiot2024phikonv2largepublicfeature} & Sep 2024 & DINOv2 & ViT-L & 1024 & $20\times$ & $\approx$ 58K  & Public \\
        Prov-GigaPath ~\citep{gigapath2024} & May 2024 & DINOv2 & ViT-G & 1536 & $20\times$ & $\approx$ 170K   & Private \\
        ResNet-IN ~\citep{he2016deep} & Dec 2015 & - & ResNet-50 & 1024 & - & 1.3M (natural images)  & Public \\
        RetCCL ~\citep{wang2023retccl} & Oct 2022 & CCL & ResNet-50 & 2048 & $20\times$  & $>$ 32K  & Public  \\
        UNI ~\citep{chen2024uni} & Mar 2024 & DINOv2 & ViT-L & 1024 & $20\times$  & $\approx$ 100K  & Mixed  \\
        UNI2-h ~\citep{UNI2h2024} & Jan 2025 & DINOv2 & ViT-H/14 & 1536 & $20\times$  & $>$ 350K  & Private  \\
        Virchow ~\citep{virchow2024} & Sep 2023 & DINOv2 & ViT-H/14 & 2560 & $20\times$ & 1.5 million  & Private \\
        Virchow2 ~\citep{virchow2_2024}\hspace{4.0mm} & Aug 2024 & Modified DINOv2  & ViT-H/14 & 2560 & $5$, $10$, $20$, $40\times$ & 3.1 million  & Private \\
        \bottomrule
        \end{tabular}
    }
    \label{table:foundation_models}
\end{table*}

\subsection{Preprocessing of WSIs} 
\label{subsection:methods_preprocessing}
All WSIs in both CHIME Multiscanner and TCGA-BRCA were preprocessed using a standardised workflow. Tissue segmentation using Otsu's thresholding~\citep{otsu1979} was first performed to remove background regions, after which non-overlapping tiles of size $256 \times 256$ pixels were extracted at a standardised resolution of $0.435\mu$m/pixel (corresponding to $20\times$ equivalent magnification for a reference scanner). This ensured that input image tiles had identical spatial resolution across all scanners, regardless of their native magnification.

To ensure high data quality within the TCGA-BRCA training data used for the supervised model evaluation, image blur was estimated using the variance of Laplacian (VL) metric~\citep{pech2000diatom}, and blurry tiles ($\mathrm{VL} < 500$) were excluded. This quality control step was not applied to CHIME Multiscanner, as blur-based tile exclusion was found to vary quite substantially across scanners for some slides, which would have introduced scanner-dependent filtering and confounded the isolation of scanner effects.

\subsection{Feature Extractors}
\label{subsection:methods_feature-extractors}

In this study, we evaluate 14 tile-level feature extractors, comprising 13 PFMs and a ResNet baseline model trained on natural images. All feature extractors are utilized as frozen encoders, i.e., their weights are not updated during downstream model training. An overview of all evaluated feature extractors is provided in Table~\ref{table:foundation_models}.

\subsubsection{Vision-Only PFMs}
The majority of the evaluated PFMs are vision-only models trained using self-supervised learning on large-scale histopathology datasets. These models predominantly utilize the DINOv2~\citep{oquab2024dinov2} framework, which combines the discriminative self-distillation objective of DINO~\citep{caron2021emerging} with the masked image modeling objective of iBOT~\citep{zhou2022ibot}.

\paragraph{State-of-the-Art ViTs}
We evaluate UNI~\citep{chen2024uni}, a ViT-Large model pretrained on over 100 million tissue tiles extracted from more than 100$\thinspace$000 WSIs across 20 tissue types. In addition, we include Phikon-v2~\citep{filiot2024phikonv2largepublicfeature} (ViT-Large), H-Optimus-0~\citep{hoptimus0} (ViT-Giant) and Prov-GigaPath~\citep{gigapath2024} (ViT-Giant), which were trained on public or private datasets ranging in size from approximately 58$\thinspace$000 to 170$\thinspace$000 WSIs. We also evaluate Virchow~\citep{virchow2024}, a ViT-Huge model trained on a private dataset comprising 1.5 million WSIs.

\paragraph{Second-Generation ViTs}
Representing the latest generation of scaled state-of-the-art ViTs, both in terms of model capacity and training data volume, we include UNI2-h~\citep{UNI2h2024}, H-Optimus-1~\citep{hoptimus1} and Virchow2~\citep{virchow2_2024}. These models employ billion-parameter architectures (ViT-Huge and ViT-Giant) and are trained on datasets containing more than 350$\thinspace$000, more than 1 million, and more than 3 million WSIs, respectively.

\paragraph{Robustness-Focused Distillation}
Distinct from the large-scale models, H0-mini~\citep{filiot2025_H0-mini} is a smaller ViT-Base model obtained through knowledge distillation~\citep{hinton2015distilling} from the larger H-Optimus-0 teacher. It is explicitly designed to retain the performance of its teacher while improving computational efficiency and robustness to domain shifts.

\subsubsection{Vision-Language Models}
In contrast to vision-only models, vision-language PFMs incorporate a textual component during pretraining. These multimodal models are designed to align the visual feature space with a semantic text space, typically by leveraging image-caption pairs extracted from medical literature. We evaluate CONCH~\citep{conch2024}, a ViT-Base model trained using the CoCa framework~\citep{yu2022coca}, which combines contrastive learning with an image captioning objective. CONCH was first pretrained on 16 million image patches and subsequently trained on more than 1.1 million image-caption pairs curated from PubMed. We also evaluate its successor, CONCHv1.5~\citep{conchv1.5}, which scales the architecture to ViT-Large and initializes its weights from UNI prior to the multimodal vision-language alignment step.

\subsubsection{Early CNN-based PFMs and ImageNet Baseline}
To benchmark progress against earlier generations of computational pathology models and to evaluate scanner robustness across different architectural paradigms, we include RetCCL~\citep{wang2023retccl} and CTransPath~\citep{Wang2023ctranspath}. RetCCL employs a ResNet-50 architecture trained using a clustering-guided contrastive learning objective, while CTransPath uses a hybrid CNN-Transformer architecture based on Swin-T~\citep{liu2021swin}. Both models were trained on public datasets that are substantially smaller than those used for state-of-the-art PFMs. Finally, ResNet-IN~\citep{he2016deep}, a standard ResNet-50 trained on the ImageNet dataset~\citep{ImageNet} of natural images, is also included as a baseline to quantify the benefit of domain-specific pretraining.

\section{Evaluation Framework}
\label{section:evaluation_framework}

To comprehensively assess the robustness of PFM feature extractors to scanner-induced domain shifts, we adopt a multi-faceted evaluation framework comprising four complementary main components: (1) a \textbf{qualitative assessment of latent feature-space geometry} using UMAP visualisations, (2) a \textbf{quantitative characterisation of embedding invariance} using distance-based metrics computed on slide-level features, (3) a \textbf{supervised benchmark of downstream clinicopathological prediction tasks} using attention-based multiple instance learning (ABMIL) models, and (4) an \textbf{evaluation of prediction consistency and calibration stability} across different scanner devices. This framework ensures that we assess not only the intrinsic geometric properties of learned representations, but also their practical reliability in scenarios that are resembling real clinical deployment.

\subsection{Unsupervised Embedding Analysis}
To evaluate the robustness to scanner-induced domain shifts independently of any specific downstream task, we conduct both qualitative and quantitative analyses of the PFM feature extractor embedding space. Our analysis operates at two complementary levels: tile-level embeddings which capture the distributional properties of individual tissue patches, and slide-level embeddings (computed by mean-pooling tile-level features) which represent entire WSIs.

\subsubsection{Qualitative Analysis via UMAP Visualisations}
We employ UMAP~\citep{mcinnes2018umap} to generate low-dimensional visualisations of the high-dimensional embedding spaces, which reveal whether scanner-specific signatures appear as distinct clustering patterns. Specifically, we visualise the first two UMAP dimensions for both tile-level (Figure~\ref{fig:umap_tile}) and slide-level embeddings (Figure~\ref{fig:umap_slide}), with each point coloured according to the scanner device used for acquisition. A high degree of colour mixing indicates scanner-invariant representations, whereas distinct clustering or separation by colour (scanner) suggests sensitivity to scanner-induced variability. While these visualisations are inherently qualitative and subject to distortions introduced by non-linear dimensionality reduction, they provide intuitive insights into scanner effects and serve as a motivating precursor to the quantitative analyses described below.

\subsubsection{Quantitative Geometric Embedding Analysis}
\label{subsection:quantative-feature-eval}
To quantitatively assess robustness to scanner-induced variability, we compute five complementary distance-based metrics in the original high-dimensional embedding space. Each metric captures a distinct aspect of embedding invariance, enabling a comprehensive characterisation of how well feature extractors preserve morphological relationships across scanner domains. 

\paragraph{Notation}
For a given feature extractor $f_\theta$, we extract a $d$-dimensional feature vector $f_\theta(t) \in \mathbb{R}^d$ from each image tile $t$. For a WSI acquired from patient  $p \in \mathcal{P}$ using scanner $s \in \mathcal{S}$, we compute a slide-level embedding $\mathbf{h}_{p,s} \in \mathbb{R}^d$ by mean-pooling over the $K$ tissue tiles $\{t_{p,s}^{(k)}\}_{k=1}^K$ extracted from that WSI:
\begin{equation}
    \mathbf{h}_{p,s} = \frac{1}{K} \sum_{k=1}^{K} f_\theta(t_{p,s}^{(k)}).
\end{equation}
This aggregation strategy yields a robust slide-level representation that is well-suited for geometric analysis of embeddings across scanner domains.

\paragraph{Metric 1: Average Pairwise Cosine Distance}
To assess the consistency of slide-level representations across scanner pairs, we compute the \textbf{Average Pairwise Cosine Distance} ($\mathcal{D}_{\cos}$) between embeddings of the same physical slides acquired using two different scanners $s_i, s_j \in \mathcal{S}$:
\begin{equation}
    \mathcal{D}_{\cos}(s_i, s_j) = \frac{1}{N} \sum_{p=1}^{N} \left( 1 - \frac{\mathbf{h}_{p,s_i} \cdot \mathbf{h}_{p,s_j}}{\|\mathbf{h}_{p,s_i}\|_2 \|\mathbf{h}_{p,s_j}\|_2} \right),
    \label{eq:cosine_dist}
\end{equation}
where $N$ is the number of patients ($N = 384$), and $\|\cdot\|_2$ denotes the $L^2$ Euclidean norm. Lower values of $\mathcal{D}_{\cos}$ indicate that slide-level embeddings corresponding to the same patient remain well-aligned across different scanners, reflecting stronger scanner invariance. Conversely, higher values suggest that scanner-induced variability introduces substantial shifts in the embedding space. This metric provides a direct and interpretable measure of embedding stability. $\mathcal{D}_{\cos}$ results, for the 14 evaluated feature extractors across all ten scanner pairs, are visualised as a heatmap in Figure~\ref{fig:consistency}A.

\paragraph{Metric 2: 1-Nearest Neighbour Match Rate}
While $\mathcal{D}_{\cos}$ captures absolute differences between slide-level embeddings, we also evaluate whether the relative ordering and local geometric relationships between slides are preserved across scanners via a cross-scanner instance retrieval task. Specifically, for each patient $p$ scanned on device $s_i$, we identify its nearest neighbour $\hat{p}$ in the embedding space corresponding to scanner $s_j$ and evaluate whether it corresponds to the same physical slide. The \textbf{1-Nearest Neighbour (1-NN) Match Rate} ($\mathcal{MR}_{1NN}$) quantifies this consistency:
\begin{equation}
    \mathcal{MR}_{1NN}(s_i, s_j) = \frac{1}{N} \sum_{p=1}^{N} \mathbb{I}\{\hat{p} = p\},
    \label{eq:1NN_match_rate}
\end{equation}
where $\mathbb{I}\{\cdot\}$ is the indicator function. Values of $\mathcal{MR}_{1NN}$ close to $100\%$ mean that, for nearly all patients, the nearest neighbour retrieved across scanners corresponds to the same underlying slide, implying that local neighbourhood structure in the embedding space is preserved across scanners and reflecting strong cross-device consistency. In contrast, lower values suggest scanner-dependent distortions of the learned feature embedding space. Results for all feature extractors and scanner pairs are reported in Figure~\ref{fig:consistency}B.

\paragraph{Metric 3: Mantel Correlation}
As $\mathcal{MR}_{1NN}$ focuses on individual slide-to-slide correspondences, we additionally assess global structural preservation of the embedding space across scanners using the \textbf{Mantel test}, which quantifies the similarity between pairwise distance matrices. For each scanner $s$, we construct a patient-to-patient distance matrix $\mathbf{M}^{(s)} \in \mathbb{R}^{N \times N}$, in which each element $\mathbf{M}_{p,q}^{(s)}$ represents the cosine distance between the slide-level embeddings of two patients $p, q$:
\begin{equation}
    \mathbf{M}_{p,q}^{(s)} = 1 - \frac{\mathbf{h}_{p,s} \cdot \mathbf{h}_{q,s}}{\|\mathbf{h}_{p,s}\|_2 \|\mathbf{h}_{q,s}\|_2}.
    \label{eq:mantel_matrix}
\end{equation}
We then compute the \textbf{Mantel correlation coefficient} $r_M$ between the distance matrices corresponding to two scanners $s_i$ and $s_j$:
\begin{equation}
    r_\mathbf{M}(s_i, s_j) = \operatorname{Corr}\left( \mathbf{M}^{(s_i)}, \mathbf{M}^{(s_j)} \right).
    \label{eq:mantel_corr}
\end{equation}
Higher values of $r_M$ indicate stronger preservation of global geometric structure across scanners, such that pairs of slides that are similar under one acquisition device remain similar under another. This metric is particularly relevant for unsupervised discovery tasks such as tissue or tumour subtyping, where clustering and manifold-based analyses depend critically on stable, scanner-invariant distance relationships. The Mantel correlation results are visualised in Figure~\ref{fig:consistency}C.

\paragraph{Metric 4: Mean Intra-Scanner Distance}
Beyond cross-scanner comparisons, we analyse the internal geometry and density of each scanner-specific embedding space by computing the \textbf{Mean Intra-Scanner Distance} for every patient. Specifically, for each patient $p$ acquired on scanner $s$, we compute the average cosine distance to all other patients $q$ in the dataset:
\begin{equation}
    \bar{d}_p^{(s)} = \frac{1}{N-1} \sum_{q \neq p} M_{p,q}^{(s)}.
    \label{eq:mean_intra_scanner_dist}
\end{equation}
We then examine the distribution of these values $\{\bar{d}_p^{(s)}\}_{p=1}^N$ across all scanners, as visualised in Figure~\ref{fig:consistency}D. This metric characterises the overall compactness and spread of the embedding space associated with each scanner. Systematic shifts in these distributions indicate scanner-dependent changes in global embedding geometry, which may impact downstream tasks that rely on distance-based representations. Importantly, this analysis also helps distinguish feature extractors with genuinely robust embeddings from those exhibiting collapsed or low-variance representations, where uniformly small distances may reflect limited discriminative capacity rather than actual scanner invariance.

\paragraph{Metric 5: Intersection-over-K Neighbourhood Consistency}
Finally, to assess the stability of local neighbourhood structures across all scanners simultaneously, we compute the \textbf{Intersection-over-K (IoK)} metric. For a given neighbourhood size $K$, let $\mathcal{N}_K^{(s)}(p)$ denote the set of the $K$ nearest neighbours of patient $p$ in the embedding space corresponding to scanner $s$. The IoK metric quantifies the proportion of neighbours that are shared across all scanners:
\begin{equation}
    \text{IoK}_K(\mathcal{S}) = \frac{1}{N} \sum_{p=1}^{N} \frac{\left| \bigcap_{s \in \mathcal{S}} \mathcal{N}_K^{(s)}(p) \right|}{K}.
    \label{eq:iok}
\end{equation}
We evaluate this metric across a range of $K$ values from $1$ to $N-1$, in order to analyse structural consistency at both local (small $K$) and global (large $K$) scales. High IoK values indicate that the local morphological neighbourhoods around each patient are preserved across scanners, meaning that slides with similar histological features remain grouped together regardless of acquisition device. This property is particularly important for applications involving similarity search, case-based reasoning, and other methods that rely on stable local neighbourhood structure. Results for the IoK metric are shown in Figure~\ref{fig:consistency}E1 for $K \in [1, 10]$, and in Figure~\ref{fig:consistency}E2 for the full range of $K$ values.

\paragraph{Summary of Embedding Robustness Metrics}
Taken together, Metric 1 – 5 provide a comprehensive assessment of scanner robustness in the learned embedding spaces. Metric~1~\&~2 evaluate cross-scanner consistency at the slide level, capturing both absolute alignment of representations and preservation of local neighbourhood relationships. Metric~3 extends this analysis to the global scale by assessing whether pairwise patient similarities are consistently maintained across scanners. Metric~4 characterises the internal geometry of scanner-specific embedding spaces, enabling detection of scanner-induced changes in compactness as well as collapsed or low-variance representations. Finally, Metric~5 assesses neighbourhood stability across all scanners over multiple scales, providing a stringent test of both local and global structural consistency. Collectively, these metrics distinguish true scanner-invariant representations from artefactual invariance or collapsed feature spaces, with direct relevance for downstream clinical tasks.

\subsection{Supervised Downstream Model Evaluation}
To assess how scanner-induced domain shifts affect downstream performance across clinicopathological prediction tasks, we benchmark all feature extractors on five clinically relevant breast cancer classification tasks: ER, PR, and HER2 status prediction from H\&E WSIs, binary Nottingham histological grade (NHG~1 vs NHG~3), and multiclass NHG (1 vs 2 vs 3). For all tasks, models are trained on TCGA-BRCA and evaluated on the CHIME Multiscanner dataset, enabling systematic assessment across five different scanner devices for the same set of patients.

As scanner variability may influence downstream prediction behaviour through multiple mechanisms, we structure the evaluation into three complementary components: \textbf{(1) predictive performance}, assessing whether discriminative ability varies across scanners;\textbf{ (2) prediction consistency}, measuring whether discrete class assignments remain consistent across scanners; and \textbf{(3) calibration stability}, examining whether predicted probabilities shift systematically even when ranking performance is preserved. Together, these components allow us to distinguish domain shifts that degrade overall performance, alter clinical decisions, or compromise probability calibration, each with distinct implications for reliable clinical deployment.

\subsubsection{ABMIL Model Architecture and Training}
For all five downstream tasks, we use an attention-based multiple instance learning (ABMIL) model~\citep{ilse2018attention} to aggregate tile-level embeddings into slide-level predictions. The ABMIL model is implemented using the CLAM framework (CLAM-SB)~\citep{lu2021data}, with the instance-level clustering constraints disabled. The model utilises a learnable gated attention mechanism to assign an attention score to each tile, enabling the network to automatically prioritise diagnostically relevant regions within each WSI before aggregating them into a final slide-level representation. This representation is subsequently passed through a linear classification layer to produce the final slide-level prediction. Hyperparameters for all tasks are set according to UNI~\citep{chen2024uni} and are summarised in  Table~\ref{tab:hyperparameters} in the supplementary material. 

To account for variability arising from model training, we train each model configuration ten times using ten stratified 80/20 random splits of the TCGA-BRCA dataset, each generated with a different random seed. These data splits are identical across all feature extractors to ensure a fair comparison. All trained models are evaluated on the complete CHIME Multiscanner test set for each of the five scanner devices, yielding $10 \times 5 = 50$ evaluation results per feature extractor and task combination.

\subsubsection{Predictive Performance}
\begin{figure*}[t]
    \centering
    \includegraphics[width=0.99\textwidth]{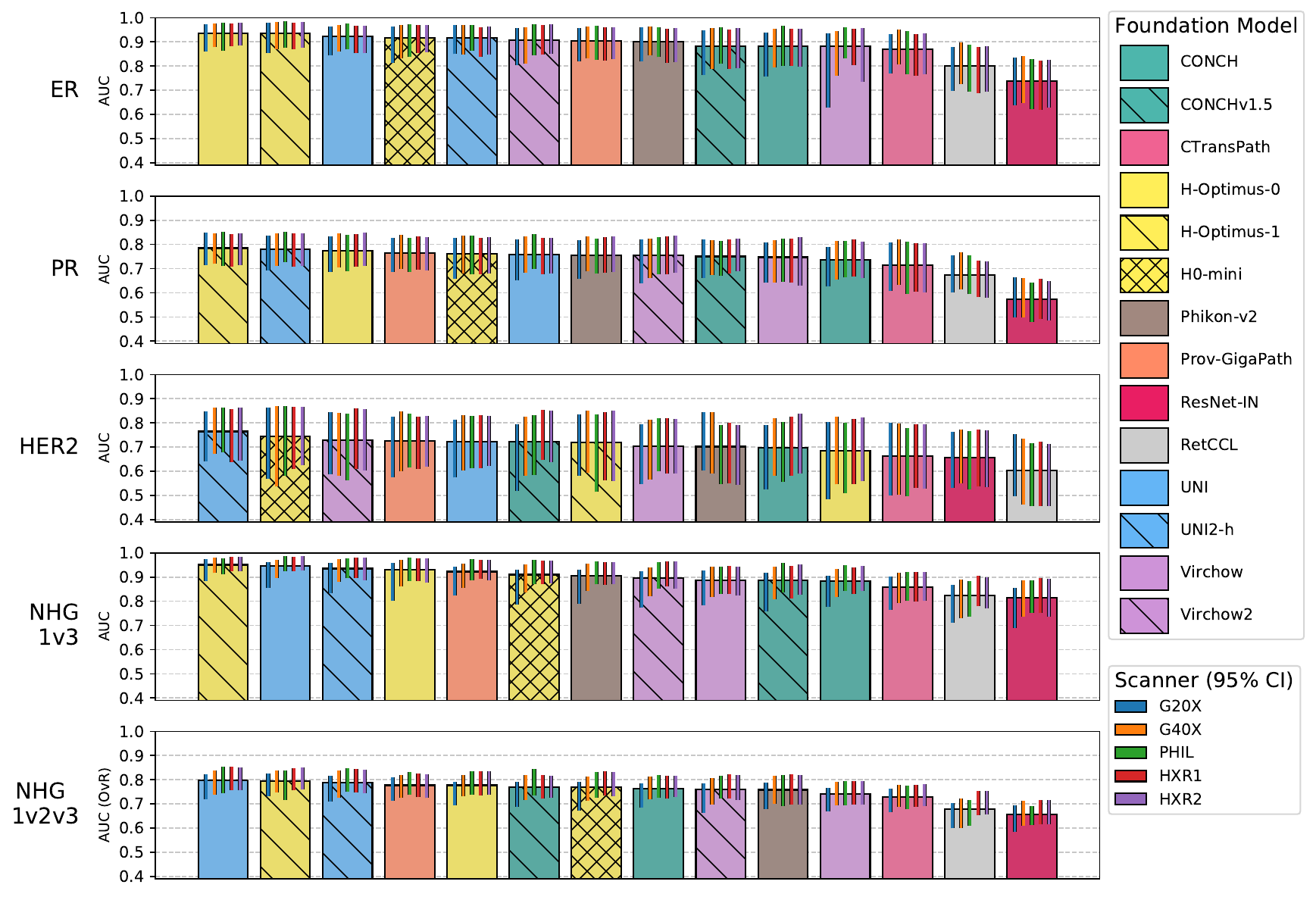}
    \caption{Results for the supervised downstream model evaluation of \emph{predictive performance}, reported in terms of AUC with 95\% bootstrap confidence intervals (1$\thinspace$000 resamples). Results are shown for each feature extractor across all five downstream clinical tasks (different rows): ER status, PR status, HER2 status, binary NHG (1 vs 3), and multiclass NHG (1 vs 2 vs 3). The main bar plots display mean AUC aggregated across all scanners, while the smaller bar plots show scanner-specific confidence intervals.}
    \label{fig:auc}
\end{figure*}

Classification performance is evaluated using the area under the receiver operating characteristic curve (AUC)~\citep{hanley1982meaning}. For binary classification tasks (ER, PR, HER2, and NHG 1 vs 3), we report the standard AUC, while for the multiclass NHG task (1 vs 2 vs 3), we report the one-vs-rest (OvR) macro-averaged AUC. To quantify uncertainty, we compute 95\% bootstrap confidence intervals using 1$\thinspace$000 resamples of the test set predictions. Figure~\ref{fig:auc} presents both the mean AUC across all scanners and scanner-specific confidence intervals for each of the five tasks.

\subsubsection{Prediction Consistency}
To quantify the consistency of downstream model predictions across different acquisition devices, we evaluate inter-scanner reliability using Fleiss' Kappa ($\kappa$)~\citep{fleiss1969large}. Specifically, we treat the five scanner devices as distinct ``raters'' that independently assign a predicted class label to each patient slide. Consequently, for every patient, the model generates five potentially different diagnostic decisions based on the input images acquired from each scanner. Fleiss' $\kappa$ measures the degree of agreement among these scanner-specific predictions beyond what would be expected by chance alone.

\begin{figure*}[t]
    \centering
    \includegraphics[width=0.99\textwidth]{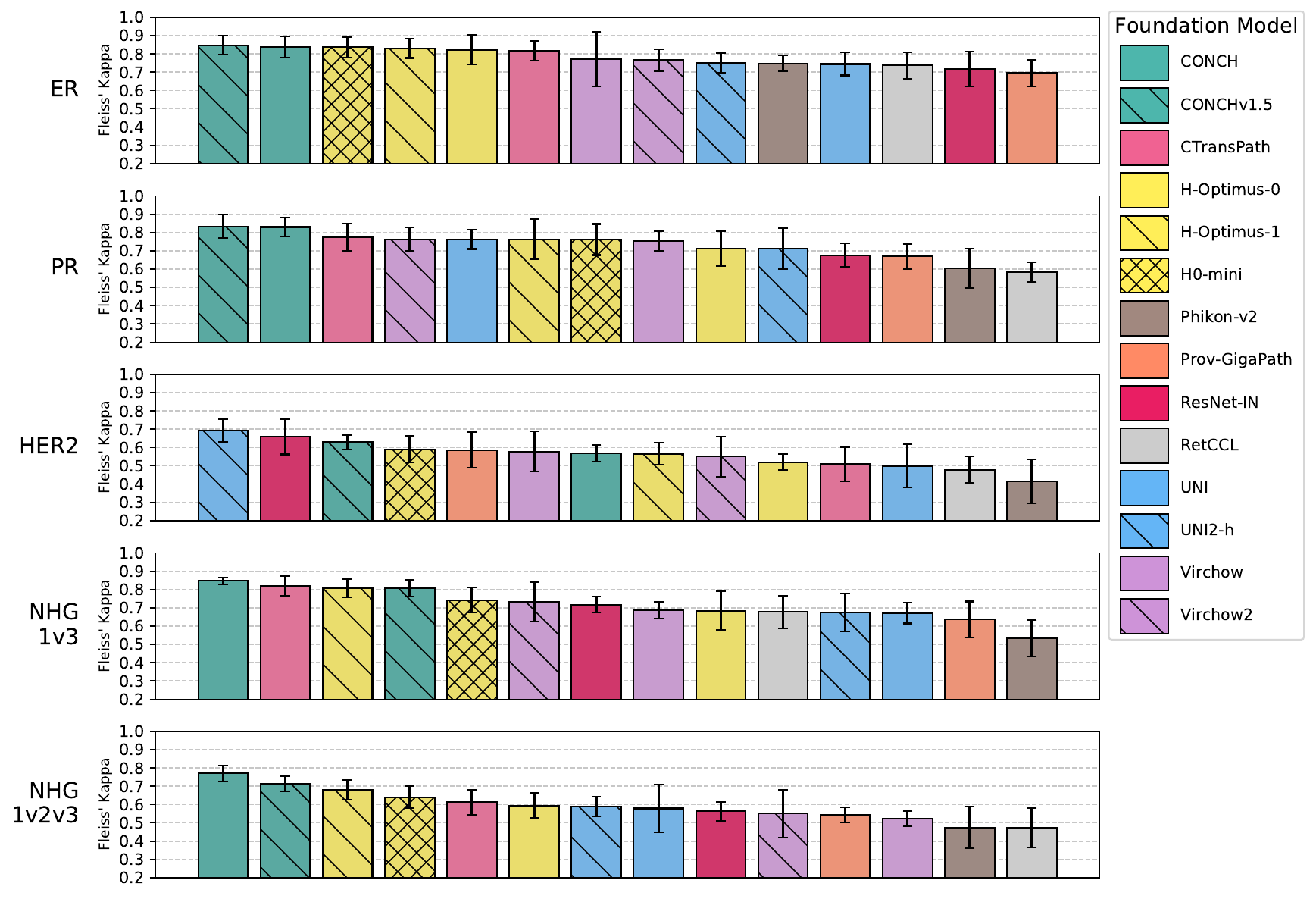}
    \caption{Results for the supervised downstream model evaluation of \emph{prediction consistency}, reported in terms of Fleiss' Kappa ($\kappa$). Results are shown for each feature extractor across all five downstream clinical tasks (different rows). The plots quantify how consistently each feature extractor produces the same classification decision for a given patient across the different scanner devices. Bars report the mean Fleiss' $\kappa$ across ten random seeds, with error bars representing the standard deviation. Higher values correspond to greater robustness.}
    \label{fig:fleiss_kappa}
\end{figure*}

This metric therefore allows us to assess the stability of the model's decision boundary across scanner domains, independently of predictive accuracy with respect to ground truth. Fleiss' $\kappa$ values are interpreted as follows:
\begin{itemize}
    \item $\kappa=1$: Perfect agreement. The model produces identical predictions for a patient regardless of the scanner used.
    \item $\kappa \leq 0$: Poor agreement. The consistency across scanners is equivalent to or worse than random chance.
\end{itemize}
We compute $\kappa$ separately for each feature extractor and task, reporting the mean and standard deviation across the ten random seeds. High $\kappa$ values indicate that a model's discrete classification decisions are robust to scanner variability, whereas lower values suggest that the choice of scanner device frequently alters the diagnostic outcome. Results are visualised in Figure~\ref{fig:fleiss_kappa}.

\begin{figure*}[t]
    \centering
    \includegraphics[width=1.0\textwidth]{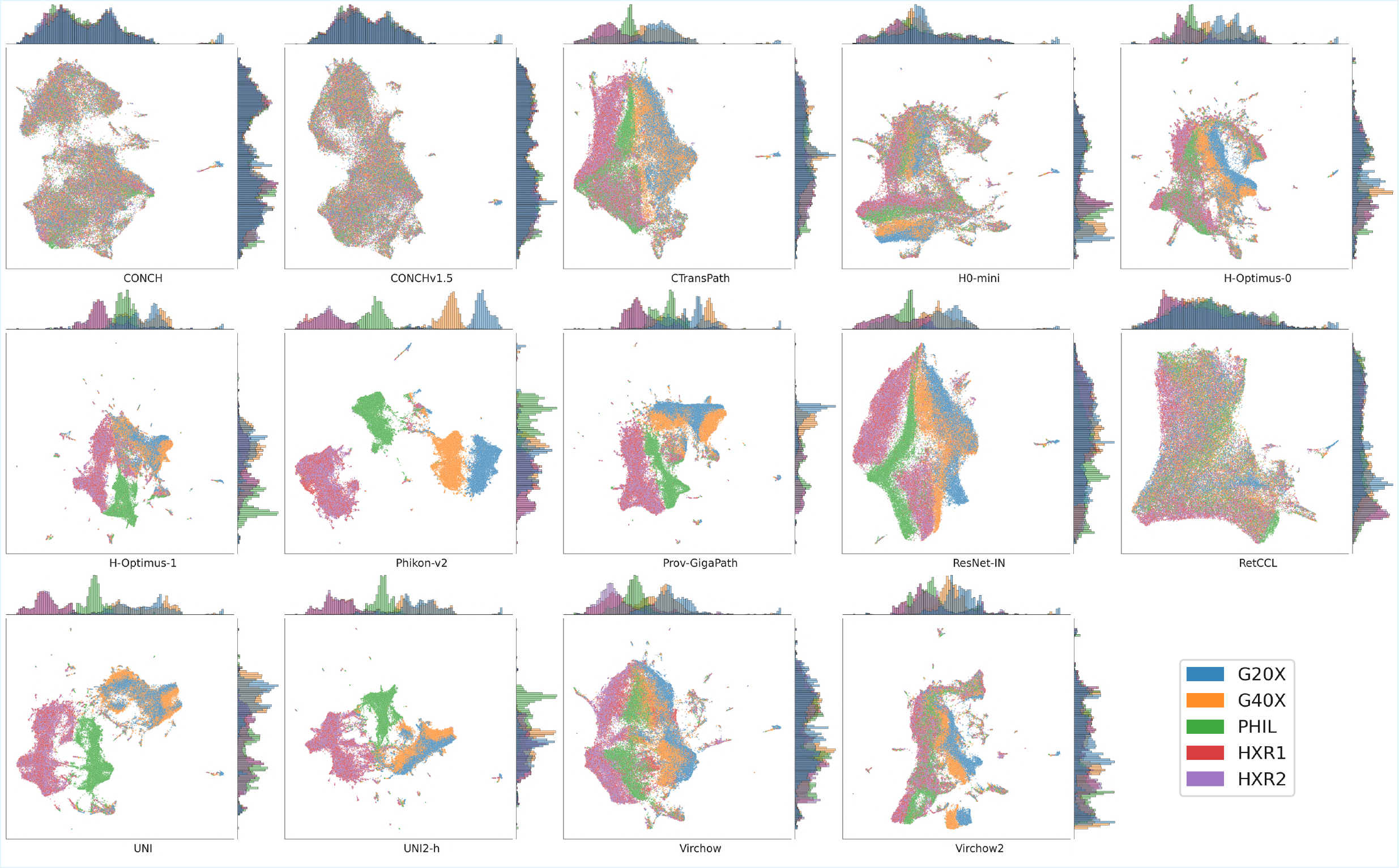}
    \caption{Low-dimensional visualisation of \emph{tile}-level embeddings using UMAP for all evaluated feature extractors. For each of the 1920 WSIs in CHIME Multiscanner (384 patients $\times$ 5 scanners $=$ 1920 WSIs), 35 tiles were randomly sampled, projected into a shared two-dimensional embedding space, and coloured according to the scanner device used for acquisition. Each subplot corresponds to a single feature extractor, and marginal plots along the x- and y-axes show the empirical density distributions for each scanner. Distinct clustering or colour separation indicates sensitivity to scanner-induced variability, whereas intermixed distributions suggest greater scanner invariance.}
    \label{fig:umap_tile}
\end{figure*}

\begin{figure*}[t]
    \centering
    \includegraphics[width=1.0\textwidth]{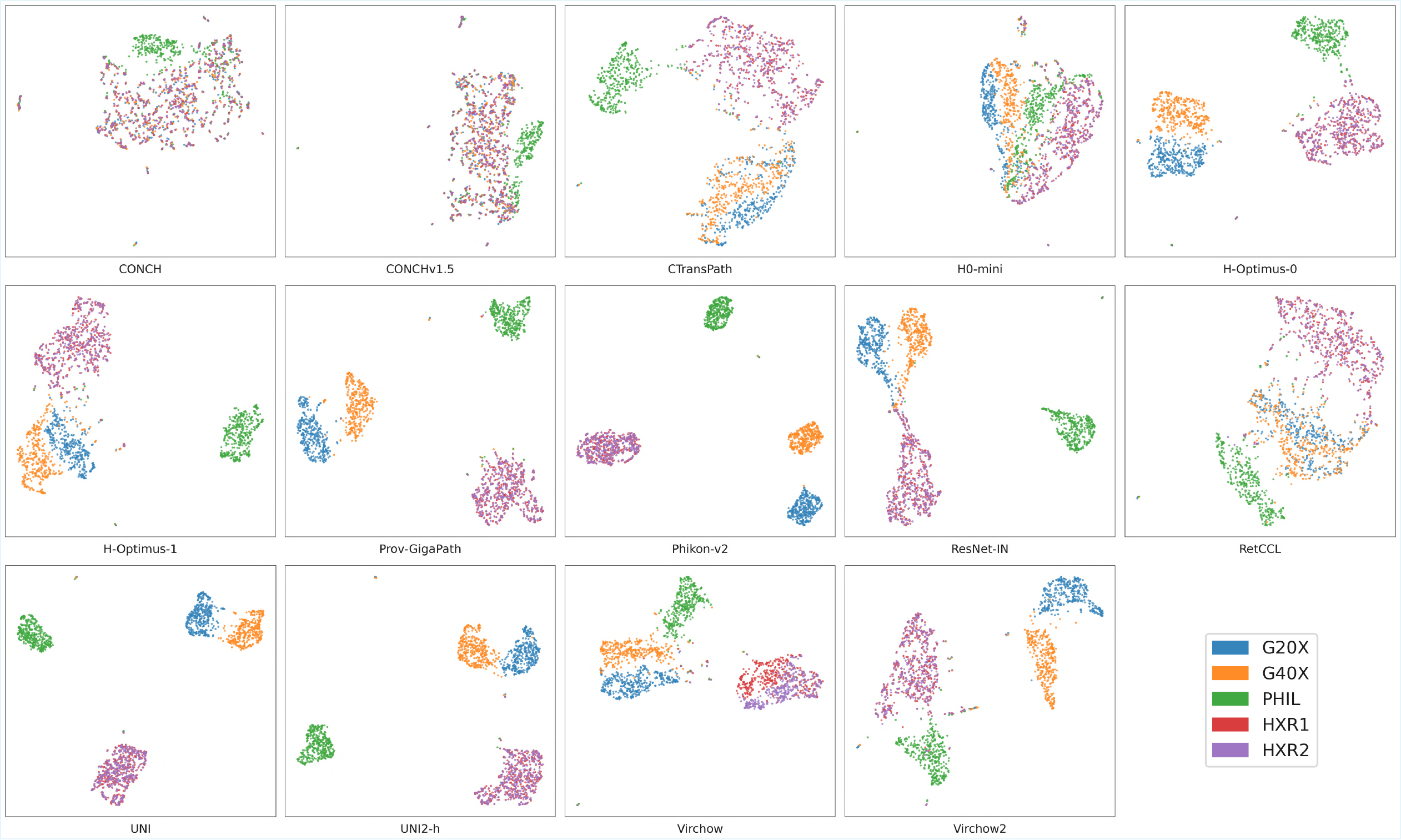}
    \caption{Low-dimensional visualisation of \emph{slide}-level embeddings using UMAP for all evaluated feature extractors. Slide-level embeddings were obtained by mean-pooling all tile-level embeddings, for each WSI in the CHIME Multiscanner dataset. Each subplot corresponds to a single feature extractor, with points coloured according to the scanner device used for acquisition. Distinct clustering or colour separation indicates sensitivity to scanner-induced variability, whereas greater overlap suggests increased scanner invariance.}
    \label{fig:umap_slide}
\end{figure*}

\begin{figure*}[t]
    \centering
    \includegraphics[width=1.0\textwidth]{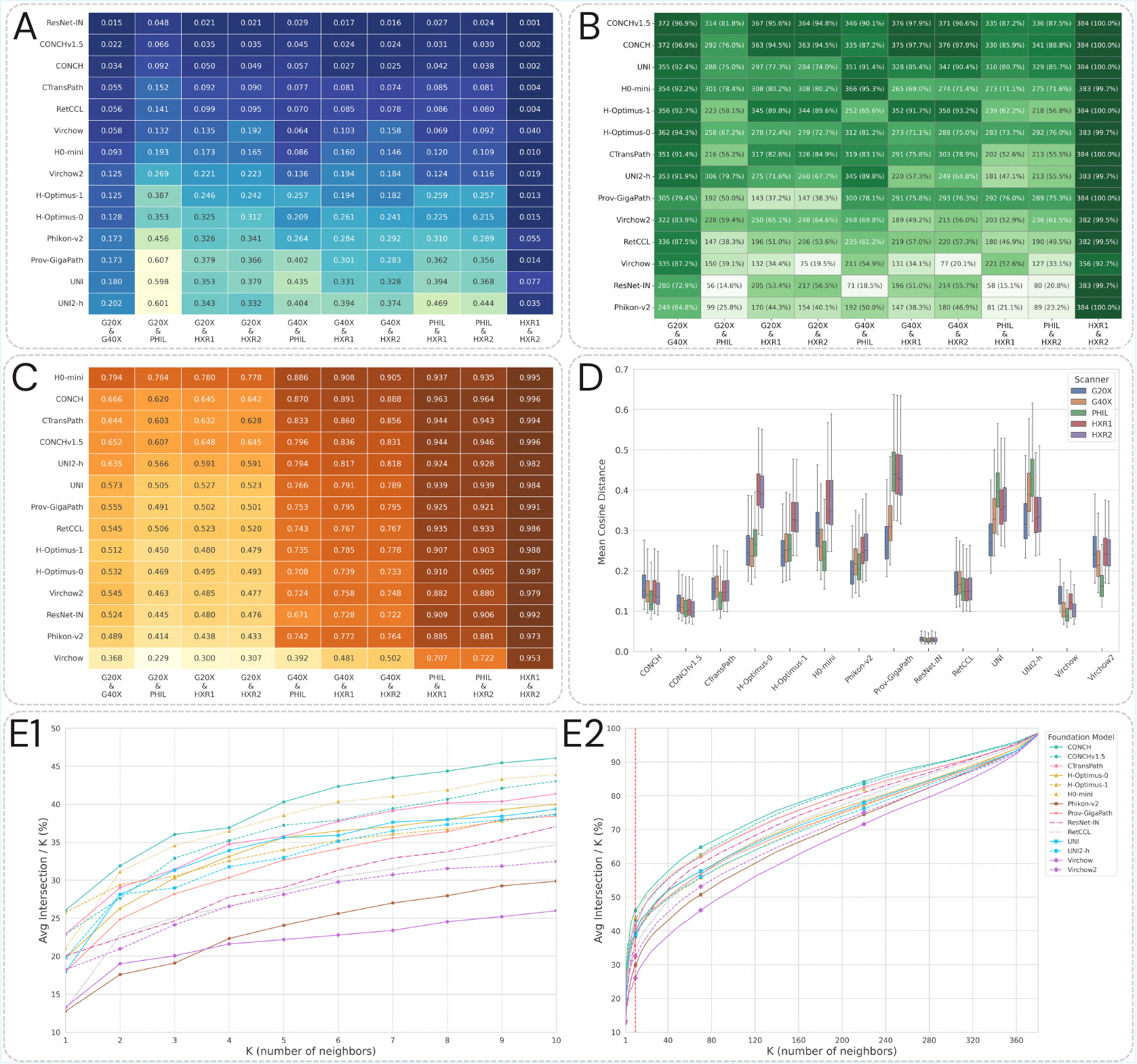}
    \caption{Quantitative geometric embedding analysis for all evaluated feature extractors, using the five complementary metrics defined in Section~\ref{subsection:quantative-feature-eval}. \textbf{A}: Heatmap of the \textbf{Average Pairwise Cosine Distance} ($\mathcal{D}_{\cos}$; Eq.~\ref{eq:cosine_dist}) between slide-level embeddings of corresponding cases acquired on different scanners. Lower values indicate higher cross-scanner embedding consistency. \textbf{B}: Heatmap of the \textbf{1-Nearest Neighbour (1-NN) Match Rate} ($\mathcal{MR}_{1NN}$; Eq.~\ref{eq:1NN_match_rate}) between scanner pairs, showing the proportion of cases for which the nearest neighbour across scanners corresponds to the same underlying slide. Higher values indicate stronger preservation of local neighbourhood structure. \textbf{C}: Heatmap of \textbf{Mantel correlation} coefficients ($r_M$; Eq.~\ref{eq:mantel_corr}) quantifying agreement between scanner-specific pairwise distance matrices. Higher values indicate stronger preservation of global embedding geometry across scanners. In panels \textbf{A}-\textbf{C}, feature extractors are ordered according to their average metric value across all scanner pairs, with the most robust models at the top. \textbf{D}: Distributions of the \textbf{Mean Intra-Scanner Distance} ($\bar{d}_p^{(s)}$; Eq.~\ref{eq:mean_intra_scanner_dist}), computed for each patient relative to all other patients within the same scanner embedding space. Shifts in these distributions indicate scanner-dependent changes in feature space geometry and density. \textbf{E1 \& E2}: \textbf{Intersection-over-K} neighbourhood consistency ($\text{IoK}_K(\mathcal{S})$; Eq.~\ref{eq:iok}) across all five scanners simultaneously. The left panel (E1) shows $K \in [1, 10]$, while the right panel (E2) shows the full range of $K$ values. Higher values indicate greater preservation of neighbourhood structure across scanners.}
    \label{fig:consistency}
\end{figure*}

\subsubsection{Calibration Stability}
Even when discriminative performance in terms of AUC is high, scanner variability may still induce bias in predicted probabilities. Such calibration instability is important, especially in a clinical deployment context, because if the same slide receives different probability estimates depending on the scanner, then a fixed decision threshold (e.g., $p > 0.5$ for a positive or high-risk classification) may yield different sensitivity and specificity across scanner devices.

To assess calibration stability, we perform pairwise comparisons of slide-level predicted probabilities across scanner pairs. For each scanner pair $(s_i, s_j)$, we construct a scatter plot in which each point corresponds to a single physical slide, with the predicted probability from scanner $s_i$ on the x-axis and from scanner $s_j$ on the y-axis. In an ideally scanner-invariant model, all points would thus lie along the diagonal line $y = x$.

To quantify systematic deviations from this diagonal, we fit locally weighted scatter plot smoothing (LOWESS) curves~\citep{Cleveland1979} to visualise the relationship between scanner pairs. LOWESS is a non-parametric regression method that fits a smooth curve to the data without assuming a specific functional form, making it suitable for visualising potential scanner-dependent calibration shifts. LOWESS curves are estimated across all ten random seeds for each feature extractor and task. For robust estimation, we adopt the following procedure. For each random seed, we bootstrap 100 LOWESS curves by randomly sampling 50\% of the slides. We then aggregate results across all seeds to compute a mean LOWESS curve with 95\% confidence intervals for each feature extractor and scanner pair. Deviations of the mean curve from the diagonal indicate systematic calibration bias, while the width of the confidence interval reflects uncertainty in the estimated relationship.

For binary classification tasks, scatter plots and LOWESS curves are constructed using the predicted probability of the positive class. For the multiclass NHG task (1 vs 2 vs 3), analyses are based on the predicted probability of NHG 3. Representative scatter plots for two selected PFMs, using the same random seed and data split, are shown in Figure~\ref{scatter_conch_h0mini_comparison}, while aggregated LOWESS curves across all ten seeds for all feature extractors are presented in Figure~\ref{fig:lowess_grade3}.


\section{Results}
\label{results}
We systematically evaluate scanner-variability robustness of all 14 feature extractors using the framework described in Section~\ref{section:evaluation_framework}. Results are organised by analysis type. We first present unsupervised embedding analyses in Section~\ref{subsec:unsupervised_results}, including qualitative visual assessment via UMAP (Section~\ref{subsubsec:qualitative_results}) and quantitative geometric analysis of the embedding space (Section~\ref{subsubsec:quantitative_results}). We then report supervised downstream model evaluation results in Section~\ref{subsec:supervised_results}, covering predictive performance (Section~\ref{subsubsec:predictive_performance}), prediction consistency (Section~\ref{subsubsec:prediction_consistency}), and calibration stability (Section~\ref{subsubsec:calibration_stability}).

\subsection{Unsupervised Embedding Analysis}
\label{subsec:unsupervised_results}

\subsubsection{Qualitative Analysis via UMAP Visualisations}
\label{subsubsec:qualitative_results}

To provide a general and unsupervised assessment of the PFM embedding spaces, we generated UMAP visualisations of \emph{tile}-level feature vectors for each feature extractor and scanner device in Figure~\ref{fig:umap_tile}. The marginal empirical distributions shown along the UMAP axes reveal clear multi-modal patterns for several PFMs, indicating pronounced scanner-dependent variation in the learned representations. The degree of scanner separation varies across models, with CONCH and CONCHv1.5 exhibiting the greatest intermixing between scanner domains.

UMAP visualisations of \emph{slide}-level embeddings in Figure~\ref{fig:umap_slide} reveal broadly similar trends. For most vision-only ViT-based PFMs, scanner-related variability manifests as well-separated clusters corresponding to different acquisition devices, indicating sensitivity to scanner-induced domain shifts. In contrast, the vision-language models CONCH and CONCHv1.5 demonstrate substantially greater overlap between scanner domains, suggesting a higher degree of robustness to scanner variability at the representation level.

We further observe that slides acquired using HXR1 and HXR2, which are two different devices but of the same scanner model, tend to cluster closely together across all feature extractors. A notable exception is Virchow, which exhibits subtle separation between HXR1 and HXR2 at the tile level that becomes more pronounced after aggregation into slide-level embeddings. This finding indicates that some PFMs can be sensitive even to subtle inter-device differences within the same scanner model.

\subsubsection{Quantitative Geometric Embedding Analysis}
\label{subsubsec:quantitative_results}
We next quantitatively analysed the consistency of slide-level feature embeddings across scanners. We first examine the \textbf{Average Pairwise Cosine Distance} ($\mathcal{D}_{\cos}$; Eq.~\ref{eq:cosine_dist}) between embeddings for all patients across scanner pairs in Figure~\ref{fig:consistency}A. Large variability in $\mathcal{D}_{\cos}$ across scanner pairs (columns in Figure~\ref{fig:consistency}A) indicates increased sensitivity to scanner-induced variability. Feature extractors are ordered according to their average $\mathcal{D}_{\cos}$ across all scanner pairs, with lower values (indicating greater cross-scanner alignment) shown at the top. We observe that the Resnet-IN baseline, together with CONCH and CONCHv1.5, achieves the best performance according to this metric, while UNI and UNI2-h rank at the bottom. We also note that, as expected, HXR1 and HXR2 (two devices of the same model) is the scanner pair that consistently exhibits the lowest $\mathcal{D}_{\cos}$ across all feature extractors.

We then evaluated the \textbf{1-Nearest Neighbour (1-NN) Match Rate} ($\mathcal{MR}_{1NN}$; Eq.~\ref{eq:1NN_match_rate}) in Figure~\ref{fig:consistency}B, which captures preservation of local neighbourhood structure across scanners. This metric exhibits substantial variability both across scanner pairs and across feature extractors. For example, for the G20X-PHIL scanner pair, CONCHv1.5 achieves an $\mathcal{MR}_{1NN}$ of $81.8\%$, whereas Phikon-v2 attains only $25.8\%$. CONCH and CONCHv1.5 consistently rank among the top-performing feature extractors across all scanner pairs, achieving the highest average $\mathcal{MR}_{1NN}$, while Phikon-v2 ranks lowest overall. These results thus highlight pronounced differences in scanner robustness among PFMs. As observed for $\mathcal{D}_{\cos}$ above, the HXR1-HXR2 scanner pair again stands out, with near-perfect $\mathcal{MR}_{1NN}$ achieved across all feature extractors.

To further assess global structural consistency of the slide-level embeddings across scanner devices, we compute the \textbf{Mantel correlation} between scanner-specific distance matrices ($r_M(s_i, s_j)$; Eq.~\ref{eq:mantel_corr}) for each feature extractor in Figure~\ref{fig:consistency}C. We observe substantial variability in Mantel correlation across both scanner pairs and feature extractors, indicating marked differences in the preservation of global embedding geometry. In particular, H0-mini achieves the highest overall Mantel correlation, while Virchow exhibits the lowest consistency across scanner pairs. CONCH and CONCHv1.5 also rank among the top-performing models. As a positive control, the HXR1-HXR2 scanner pair once again demonstrates high consistency across all feature extractors.

When analysing the \textbf{Mean Intra-Scanner Distance} ($\bar{d}_p^{(s)}$; Eq.~\ref{eq:mean_intra_scanner_dist}) results in Figure~\ref{fig:consistency}D, we observe pronounced inter-scanner differences in the distributions of mean cosine distances for the majority of vision-only ViT-based models, including H-Optimus-0, H-Optimus-1, Prov-GigaPath, UNI, UNI2-h, Virchow and Virchow2. These distributional shifts indicate that the choice of scanner device impacts not only pairwise relationships between slides, but also the global geometry and density of the feature space. Moreover, we observe a consistent correspondence between scanner pairs exhibiting the highest pairwise cosine distances in Figure~\ref{fig:consistency}A (e.g., G20X-PHIL for UNI and UNI2-h) and those showing the largest distributional shifts in mean intra-scanner distance in Figure~\ref{fig:consistency}D.

The ResNet-IN baseline exhibits the lowest overall $\bar{d}_p^{(s)}$ across all scanners in Figure~\ref{fig:consistency}D, indicating a highly compact feature space. Although ResNet-IN achieves the lowest $\mathcal{D}_{\cos}$ values in Figure~\ref{fig:consistency}A, this apparent robustness is misleading when considered in isolation, as the model simultaneously ranks near the bottom in terms of both $\mathcal{MR}_{1NN}$ (Figure~\ref{fig:consistency}B) and Mantel correlation $r_M$ (Figure~\ref{fig:consistency}C). Taken together with the distributional evidence in Figure~\ref{fig:consistency}D, these results indicate that the low $\mathcal{D}_{\cos}$ values of Resnet-IN are driven by a non-expressive, collapsed embedding space in which all samples are densely clustered, rather than by genuine scanner invariance. In contrast, CONCH and CONCHv1.5 consistently rank among the top-performing models across all three metrics ($\mathcal{D}_{\cos}$, $\mathcal{MR}_{1NN}$ and $r_M$), indicating that they maintain expressive feature representations capable of preserving meaningful morphological distinctions across scanners.

The analysis of shared neighbourhood structure using the \textbf{Intersection-over-K} ($\text{IoK}_K(\mathcal{S})$; Eq.~\ref{eq:iok}) in Figure~\ref{fig:consistency}E1~\&~E2 reveals trends consistent with the preceding four embedding metrics. In particular, CONCH, CONCHv1.5, and H0-mini consistently exhibit the highest $\text{IoK}_K(\mathcal{S})$ values across a wide range of neighbourhood sizes $K$, indicating strong preservation of local neighbourhood structure across scanners. This stability suggests that, for these models, relative similarities between patient cases remain largely invariant to the choice of acquisition device. In contrast, Virchow, Virchow2 and Phikon-v2 consistently exhibit the lowest $\text{IoK}_K(\mathcal{S})$ values across neighbourhood sizes $K$.

Taken together, the quantitative geometric analyses reveal consistent patterns in scanner robustness across feature extractors. Across all five embedding metrics, CONCH and CONCHv1.5 rank among the most robust models. H0-mini also performs favourably, often outperforming its larger H-Optimus-0 teacher, suggesting that its robustness-focused distillation strategy is effective. In contrast, several vision-only ViT-based models, including Phikon-v2, Virchow and Virchow2, exhibit substantially higher sensitivity, ranking among the weakest performers across most metrics. Notably, increased model scale or newer versions do \emph{not} uniformly improve scanner robustness: UNI2-h does not consistently outperform UNI, H-Optimus-1 does not clearly improve over H-Optimus-0, and Virchow2 shows similar or greater sensitivity compared to Virchow. We further observe that CTransPath is not systematically less robust than recent state-of-the-art ViT models and, in several cases, outperforms both UNI variants and the H-Optimus models. Overall, these results indicate that scanner robustness reflects specific architectural and training choices rather than model size, recency, or pretraining dataset scale.

\subsection{Supervised Downstream Model Evaluation}
\label{subsec:supervised_results}

\begin{figure*}[t]
    \centering
    \includegraphics[width=1.0\linewidth]{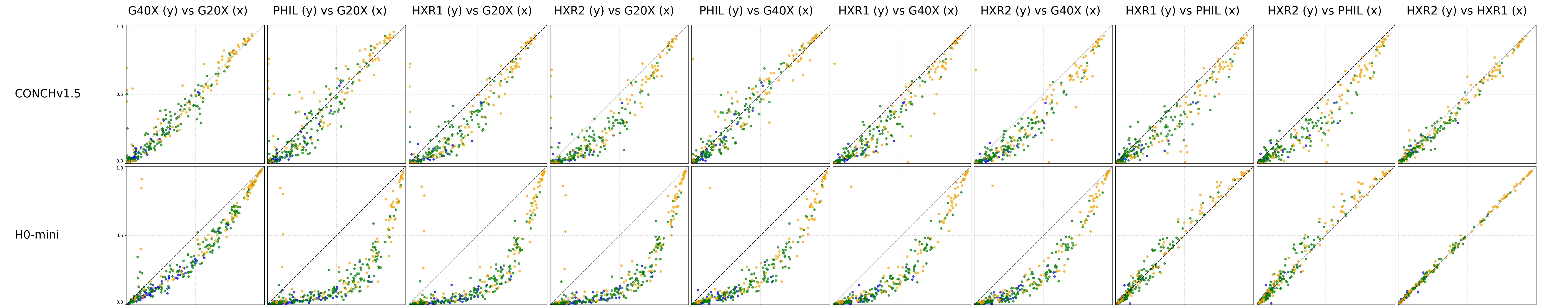}
    \caption{Results for the supervised downstream model evaluation of \emph{calibration stability}, showing representative examples of scanner-dependent calibration shifts for the multiclass NHG (1 vs 2 vs 3) task. Results are shown for CONCHv1.5 and H0-mini (rows) across all scanner pairs (columns). In each scatter plot, points correspond to individual patients, with the predicted probability from scanner $s_i$ shown on the x-axis and from scanner $s_j$ on the y-axis. Points are coloured according to the ground truth NHG label. Plots visualise the predicted probability of NHG 3 and are shown for a single representative random seed. Deviations from the diagonal line indicate scanner-dependent calibration shifts. Results for all evaluated feature extractors are shown in Figure~\ref{fig:scatter_all_fms_seed7} in the supplementary material.}
    \label{scatter_conch_h0mini_comparison}
\end{figure*}

\begin{figure*}[t]
    \centering
    \includegraphics[width=0.895\linewidth]{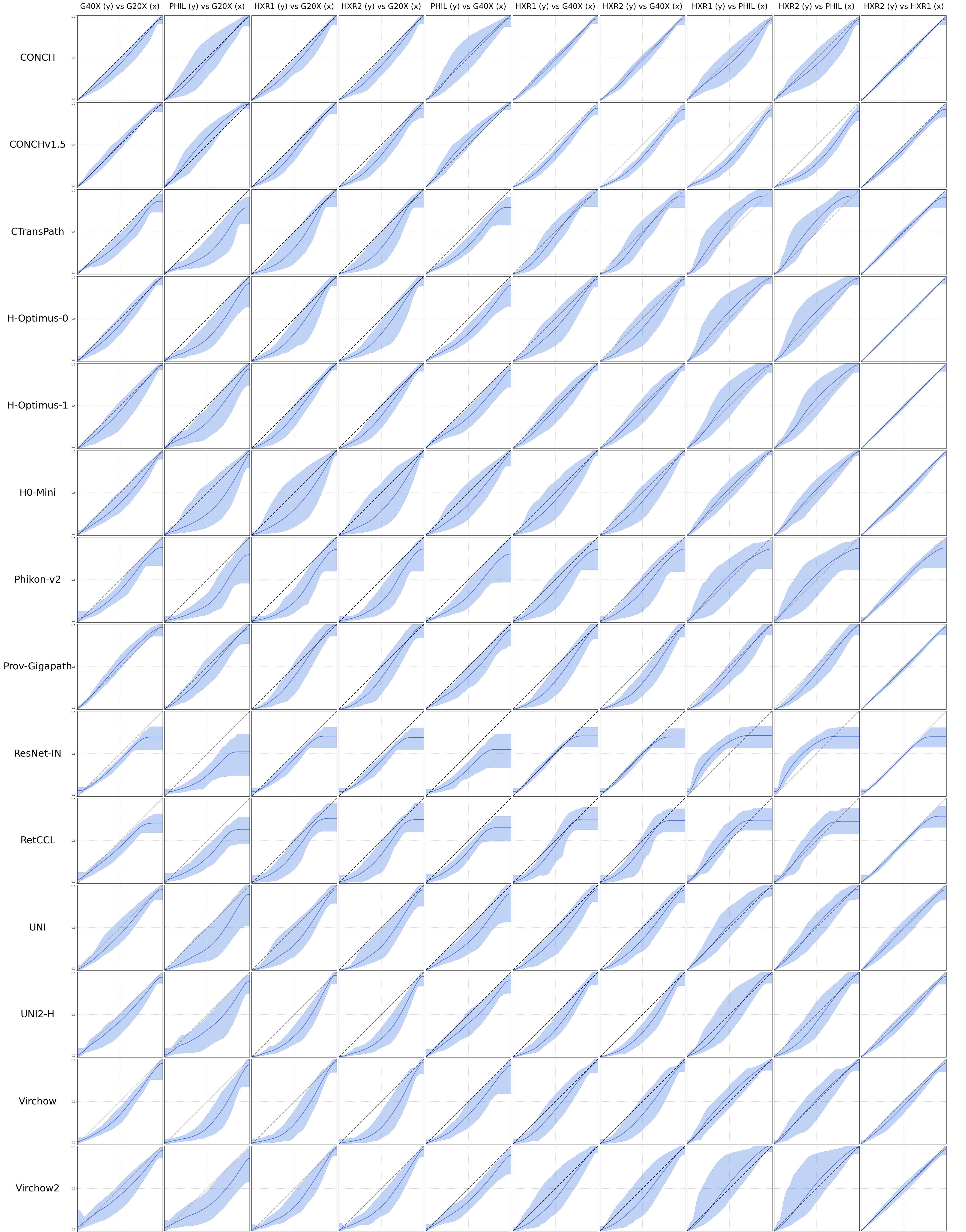}
    \caption{Results for the supervised downstream model evaluation of \emph{calibration stability}, showing LOWESS curves for the multiclass NHG (1 vs 2 vs 3) task. Results are shown for all evaluated feature extractors (rows) across all scanner pairs (columns). LOWESS curves are bootstrapped and aggregated across ten random seeds, yielding a mean curve (solid line) with 95\% confidence intervals (shaded region). Deviations of the mean LOWESS curve from the diagonal indicate systematic scanner-dependent calibration bias. Results for the four remaining downstream clinical tasks are shown in Figure~\ref{fig:lowess_er}~-~\ref{fig:lowess_grade2} in the supplementary material.}
    \label{fig:lowess_grade3}
\end{figure*}

\subsubsection{Predictive Performance}
\label{subsubsec:predictive_performance}
Next, we evaluated downstream predictive performance using task-specific supervised benchmarks to assess how scanner-induced variability affects classification accuracy. Figure~\ref{fig:auc} summarises performance in terms of AUC across all five downstream tasks for each feature extractor. Overall, performance varies across PFMs, but when considering the 95\% bootstrap confidence intervals, all ViT-based feature extractors achieve similar AUCs with some variability. We note that all estimates are within bootstrap confidence limits. Nonetheless, consistent performance trends are apparent across tasks, with the UNI and H-Optimus models achieving particularly strong performance. In contrast, the two CNN-based models RetCCL (self-supervised) and ResNet-IN (ImageNet pretrained) exhibit substantially lower performance across all tasks. For certain tasks, most notably NHG classification, we additionally observe systematic variation in predictive performance across scanner devices (e.g., consistently lower performance for G20X across most feature extractors), suggesting that scanner-induced domain shifts can impact downstream classification accuracy even when overall AUC remains high. We note that the models providing highest scores in robustness (i.e., CONCH, CONCHv1.5, and H0-mini) are not top-ranked in the supervised prediction tasks.

\subsubsection{Prediction Consistency}
\label{subsubsec:prediction_consistency}
To complement the predictive performance analysis, we evaluate how consistently each feature extractor produces the same classification decision for a given patient across different scanner devices using Fleiss' Kappa ($\kappa$) in Figure~\ref{fig:fleiss_kappa}. Across all five tasks, Fleiss' $\kappa$ values range from approximately 0.4 (Phikon-v2 on HER2) to 0.85 (CONCHv1.5 on ER), revealing substantial variation in prediction consistency both across feature extractors and across tasks.

The vision-language models CONCH and CONCHv1.5 achieve the highest inter-scanner agreement for the ER, PR, binary NHG (1 vs 3) and multiclass NHG (1 vs 2 vs 3) tasks, with consistently narrow standard deviations across the ten random seeds. This stability suggests that their superior robustness reflects intrinsic properties of the feature extractors rather than artefacts of particular training initialisations. CTransPath also emerges as a consistently strong performer, ranking among the top models for all tasks except HER2. This is noteworthy given that CTransPath is an earlier and smaller model compared to recent large-scale vision-only ViTs. In contrast, RetCCL, despite a comparable training data volume, performs poorly across most tasks. The key distinction lies in their architectures: CTransPath employs a hybrid CNN-Transformer design, whereas RetCCL relies on a pure ResNet-50 backbone, supporting the notion that architectural design choices may be more influential for scanner robustness than model scale or training data volume alone.

The distilled model H0-mini consistently outperforms its teacher H-Optimus-0 in terms of Fleiss' $\kappa$ and performs comparably to, or slightly below, H-Optimus-1. This finding aligns with the explicit robustness objectives incorporated during distillation, and mirrors the results of the unsupervised embedding analysis in Section~\ref{subsubsec:quantitative_results}, where H0-mini also demonstrates relatively high cross-scanner consistency. Large-scale vision-only ViTs (UNI, UNI2-h, H-Optimus-0, H-Optimus-1, Virchow, and Virchow2) exhibit moderate to high consistency overall but generally trail behind the vision-language models, particularly for PR and multiclass NHG. Although these vision-only ViTs achieve strong discriminative performance (Figure~\ref{fig:auc}), their discrete classification decisions thus appear more susceptible to scanner-induced shifts. Prov-GigaPath and Phikon-v2 are consistently among the weakest performers, especially for the more challenging tasks.

HER2 prediction exhibits a distinct pattern compared to the other tasks. Notably, ResNet-IN achieves relatively high inter-scanner agreement despite poor embedding robustness and low AUC performance. This apparent paradox is likely driven by the pronounced class imbalance in HER2 status (only 8.8\% HER2 positive in the CHIME Multiscanner dataset), combined with ResNet-IN’s tendency to produce low-confidence predictions clustered near the decision threshold. Under such conditions, consistently predicting the majority class can yield high agreement by chance while offering limited discriminative value, underscoring the importance of interpreting Fleiss' $\kappa$ alongside complementary metrics such as AUC rather than in isolation.

Overall, the prediction consistency results broadly mirror the trends observed in the quantitative embedding analysis (Section~\ref{subsubsec:quantitative_results}). However, supervised model training appears to partially mitigate scanner-induced variability, as evidenced by the higher-than-anticipated Fleiss' $\kappa$ values for H-Optimus-1 relative to its unsupervised geometric robustness.

\subsubsection{Calibration Stability}
\label{subsubsec:calibration_stability}
We then evaluated the stability of model predictions beyond discriminative AUC performance by analysing calibration consistency, i.e. whether predicted class probabilities remain stable across different scanner devices. 

To this end, we performed pairwise comparisons of slide-level predicted probabilities for the same physical slides scanned on different devices. Figure~\ref{scatter_conch_h0mini_comparison} illustrates representative examples of scanner-dependent calibration shifts for CONCHv1.5 and H0-mini on the multiclass NHG (1 vs 2 vs 3) task, visualising the predicted probability of NHG 3, for a single random seed. In an ideally scanner-invariant model, predictions would lie along the diagonal line $y = x$. As shown in Figure~\ref{scatter_conch_h0mini_comparison}, H0-mini exhibits a more pronounced deviation from the diagonal than CONCHv1.5 e.g. for the HXR1-G20X scanner pair, whereas both models show near-perfect alignment for HXR1-HXR2, as expected for two devices of the same scanner model. Corresponding scatter plots for all feature extractors using the same random seed are provided in Figure~\ref{fig:scatter_all_fms_seed7}.

To systematically characterise calibration behaviour across all feature extractors, scanner pairs and random seeds, we further analyse bootstrapped LOWESS curves summarising scanner-pairwise probability relationships. Results for the multiclass NHG task are shown in Figure~\ref{fig:lowess_grade3}, with corresponding analyses for the remaining tasks presented in Figure~\ref{fig:lowess_er}~-~\ref{fig:lowess_grade2}. Overall, these results indicate that predicted probabilities from supervised models are generally not well calibrated across scanner devices. For most scanner pairs, we observe systematic deviations from the diagonal, reflecting scanner-dependent calibration bias that varies in magnitude across feature extractors. Importantly, this lack of calibration stability suggests that, although many models largely preserve the relative ranking of cases across scanners, their absolute probability estimates remain sensitive to the acquisition device.

These findings have important implications for both multi-site research studies and clinical deployment. Specifically, this means that risk scores, prediction probabilities, and classification labels assigned using a fixed decision threshold, may not transfer reliably across different scanner contexts.

\paragraph{Summary of Results}
Across both unsupervised and supervised analyses, we observed consistent evidence of scanner-induced domain shifts affecting feature representations and downstream predictions. The vision-language models CONCH and CONCHv1.5 demonstrated slightly higher robustness with respect to embedding geometry, prediction consistency, and calibration stability, but did not perform at the top with respect to supervised tasks. H0-mini also showed improved robustness relative to its teacher model in some assessments, but was not robust with respect to calibration stability. In contrast, several vision-only ViT-based models exhibit greater sensitivity to scanner variability despite strong discriminative performance, and increased model size or newer versions do not uniformly confer improved robustness. Notably, while supervised training partially mitigates scanner effects on discrete predictions, systematic calibration shifts persist across most models.


\section{Discussion}
We performed a systematic, multi-metric evaluation of 14 PFM feature extractors on the CHIME Multiscanner dataset to characterise how whole-slide scanner devices impact learned representations and downstream supervised clinicopathological prediction tasks. Our analyses span qualitative UMAP visualisations, quantitative embedding geometry metrics, supervised downstream classification benchmarks, prediction consistency (Fleiss’ $\kappa$) and calibration stability, and reveals a consistent picture: despite large-scale self-supervised pretraining on diverse pan-cancer datasets, current state-of-the-art PFMs remain sensitive to scanner-induced domain shifts when other sources of variability are controlled. In particular, the choice of scanner device introduces measurable domain shifts in embedding geometry that translate into variability in downstream decisions and, critically, introduces calibration bias in supervised prediction tasks. Importantly, robustness to scanner variability does not scale monotonically with model size or recency. Instead, architectural design choices, pretraining objectives, and training data, appear to play a decisive role. Despite some variability in robustness across PFMs evaluated, none of them offers satisfactory robustness.

\subsection{Interpretation of Embedding Analysis}
At the embedding level, most vision-only ViT models exhibit pronounced scanner-dependent clustering in both tile- and slide-level representations, as well as elevated pairwise cosine distances and reduced neighbourhood consistency across scanner pairs. These findings indicate that many PFMs are sensitive to scanner-related variability in WSIs in addition to morphological features, despite extensive data augmentation during self-supervised pretraining. 

The vision-language models CONCH and CONCHv1.5 demonstrated greater scanner invariance across all embedding metrics. This is most likely due to the fact that these models were trained on substantially more diverse data: CONCH is pretrained on over one million image-caption pairs curated from figures and associated captions in PubMed. This broader and more heterogeneous visual-semantic training corpus may regularise the learned representations, reducing reliance on scanner-specific low-level cues and improving robustness to acquisition-related variability. Another possibility is that the multimodal alignment objective itself might encourage representations that prioritise semantic-related morphology over scanner-dependent patterns. Our experiments cannot disentangle the relative contributions of multimodal alignment versus data diversity, however, these factors would be of interest to characterise further in future studies.

Moreover, targeted robustness-oriented training strategies also appear beneficial. The distilled model H0-mini consistently outperforms its larger teacher model, H-Optimus-0, across multiple embedding stability metrics and downstream prediction consistency measures. This observation supports the notion that robustness-focused distillation can improve cross-scanner generalisation without increasing model size. However, despite improved performance in  many of our assessments, H0-mini had poor performance with respect to calibration. Furthermore, we noted that increased scale or newer model versions do not appear to uniformly confer improved robustness: several recent large-scale ViT models, including UNI2-h, Virchow2 and H-Optimus-1, remain sensitive to scanner-induced shifts, highlighting that robustness is not an automatic consequence of larger datasets or model capacity.

Finally, we note that slides acquired using HXR1 and HXR2, two physically different scanner devices of the same model, consistently cluster closely together across nearly all feature extractors, providing an expected positive control for scanner-induced variation. This observation confirms that the embedding analyses are sensitive to genuine acquisition differences rather than spurious noise. A notable exception is Virchow, which exhibits subtle separation between HXR1 and HXR2 at the tile level that becomes more pronounced after aggregation into slide-level embeddings. This suggests that some PFMs may capture even fine-grained inter-device differences within the same scanner model, further underscoring the sensitivity of learned representations to acquisition-specific characteristics.

\subsection{Impact on Downstream Predictions and Clinical Deployment}
While most ViT-based models achieve strong discriminative performance as measured by AUC, these results mask important failure modes. Recent state-of-the-art PFMs such as UNI2-h and H-Optimus-1 consistently achieve high AUCs across scanners and outperform earlier non-ViT models such as RetCCL and CTransPath, demonstrating that modern PFMs generalise well in terms of discriminative accuracy. However, high AUC alone does not guarantee stable decision-making or calibrated probability estimates across scanners. Systematic calibration shifts are prevalent across scanner pairs, as evidenced by LOWESS analyses, even when the relative ranking of cases is preserved. 

Consequently, risk scores, predicted class probabilities and dependent variables in regression tasks cannot be assumed to be directly comparable or well calibrated across different scanner contexts, including clinical deployment sites using different scanner devices. This also impacts the use of fixed decision thresholds that are calibrated on data from one scanner model, but is intended to be used in another context. Such discrepancies can lead to reduced performance in research studies, and pose a risk of inconsistent clinical decision-making partially impacted by scanner device. 

These findings underscore that domain generalisation in computational pathology must extend beyond discriminative accuracy to include probability calibration and decision stability across the scanners used in real-world workflows. Although supervised downstream training partially mitigates scanner effects on discrete predictions, it does not resolve calibration instability, indicating that explicit calibration-aware strategies are required for reliable clinical deployment. From a practical standpoint, these results motivate several recommendations for both model development and deployment practice. Model developers should consider incorporating robustness-oriented objectives, scanner-aware augmentation, or other measures to improve robustness, and evaluate embedding stability using geometric metrics in addition to AUC. For deployment of current generation non-robust PFMs, implementation of additional measures to mitigate and manage domain shifts due to scanner variability should be considered.

\subsection{Limitations \& Future Work}
This study has several limitations. First, our analysis focuses on a single tissue type (breast cancer resections), and scanner effects may differ across tissues, stains, or specimen preparations. Second, although we evaluated a broad selection of state-of-the-art PFMs, the rapid pace of development in the field means that newly released models may exhibit different robustness characteristics. Third, tiles were not spatially registered across scanners. Consequently, mean-pooled slide-level embeddings were used for most embedding analyses, which may obscure spatially local scanner effects or dilute region-specific sensitivity. Fourth, downstream task performance was evaluated using an ABMIL aggregation strategy, and alternative pooling or aggregation methods may interact differently with feature extractor robustness. A comprehensive evaluation of potential mitigation strategies, including a variety of different normalisation and augmentation techniques, is beyond the scope of this study and is left for future work. Here, we deliberately focus on the cleanest possible experimental setup to characterise the inherent robustness of feature extractors, in the absence of scanner-specific preprocessing.

\section{Conclusion}
This study presents a comprehensive multiscanner benchmark of 14 pathology feature extractors using the CHIME Multiscanner dataset, demonstrating that current state-of-the-art pathology foundation models are not invariant to scanner-induced domain shifts. Although most models achieve strong discriminative performance as measured by AUC, this apparent robustness masks a critical failure mode: scanner variability systematically distorts embedding geometry and downstream model calibration, leading to scanner-dependent bias that can negatively impact both research studies and introduce additional risk during use in clinical contexts.

None of the models evaluated in this study exhibited satisfactory robustness towards scanner device variability in input images. We therefore conclude that robustness to scanner variability is not a simple function of training data scale, model size, or model recency. Instead, factors such as the diversity of training data and architectural design or training strategies may play an important role. This is consistent with our observation that vision-language models exhibited comparatively improved robustness across embedding stability, prediction consistency, and probability calibration. Furthermore, robustness-oriented model distillation also offers a promising strategy for improving scanner invariance without increasing model size. 

Taken together, we conclude that none of the current generation PFMs are robust towards scanner-induced variability. Further investigations will therefore be needed to develop and evaluate modelling and training strategies that can improve PFM robustness. More broadly, our findings emphasise that achieving safe and generalisable computational pathology systems requires moving beyond accuracy-centric evaluation and scaling paradigms. Future development of pathology foundation models should explicitly prioritise robustness and calibration stability across acquisition devices, supported by comprehensive multiscanner validation. Scanner-induced domain shifts should be a central consideration when designing research studies and trials, as well in clinical deployment of solutions using pathology foundation models.

\section*{Data Availability}

The whole-slide images for the utilized TCGA-BRCA dataset are available at the GDC Data Portal \url{https://portal.gdc.cancer.gov}. The CHIME Multiscanner dataset cannot be made publicly available due to restrictions relating to sensitive patient-related information.
\section*{Code Availability}

The code for training the supervised ABMIL models in this study is based on the CLAM repository, which is available at \url{https://github.com/mahmoodlab/CLAM}. Further implementation details are available from ET upon reasonable request.

\section*{Acknowledgments}

This work was supported by funding from the Swedish Cancer Society (Cancerfonden), SeRC (Swedish e-science Research Centre, EmPHAsis), the Swedish Innovation Agency - VINNOVA (SWAIPP-2, Swedish AI Precision Pathology2), and the Swedish Research Council. The results shown here are in part based upon data generated by the TCGA Research Network: \url{https://www.cancer.gov/tcga}.
\section*{Author Contributions}

ET was responsible for preparation of figures and tables, and manuscript drafting. ET and FKG were responsible for software implementation. KLE was responsible for data processing of the CHIME Multiscanner dataset. FKG and MR were responsible for project conceptualisation and supervision. MR for funding acquisition. All authors have contributed to the design of experiments, interpretation of results and manuscript editing. 

\section*{Competing Interests}

MR is a co-founder and shareholder of Stratipath AB. KLE is employed by Stratipath AB and holds employee stock options. ET and FKG have no competing interests to declare.

\section*{Declaration of Generative AI Use}

During the preparation of this manuscript, the authors used Gemini 2.5 and ChatGPT 5.2 to assist with language editing and drafting, including identifying typographical and grammatical errors, and suggesting alternative phrasings. All content produced using these tools was critically reviewed, edited and validated by the authors, who take full responsibility for the final content of the manuscript.

{
\small
\bibliographystyle{plainnat}
\bibliography{references}
}

\clearpage
\appendix
\onecolumn

\renewcommand{\thefigure}{S\arabic{figure}}
\setcounter{figure}{0}

\renewcommand{\thetable}{S\arabic{table}}
\setcounter{table}{0}

\renewcommand{\theequation}{S\arabic{equation}}
\setcounter{equation}{0}

\subsection*{\centering{Scanner-Induced Domain Shifts Undermine the Robustness of Pathology Foundation Models}}
\section*{\centering{Supplementary Material}}

\vspace{6.0mm}

\section{Supplementary Figures}
\label{appendix:figures}

\begin{figure}[b]
    \centering
    \includegraphics[width=0.80\textwidth]{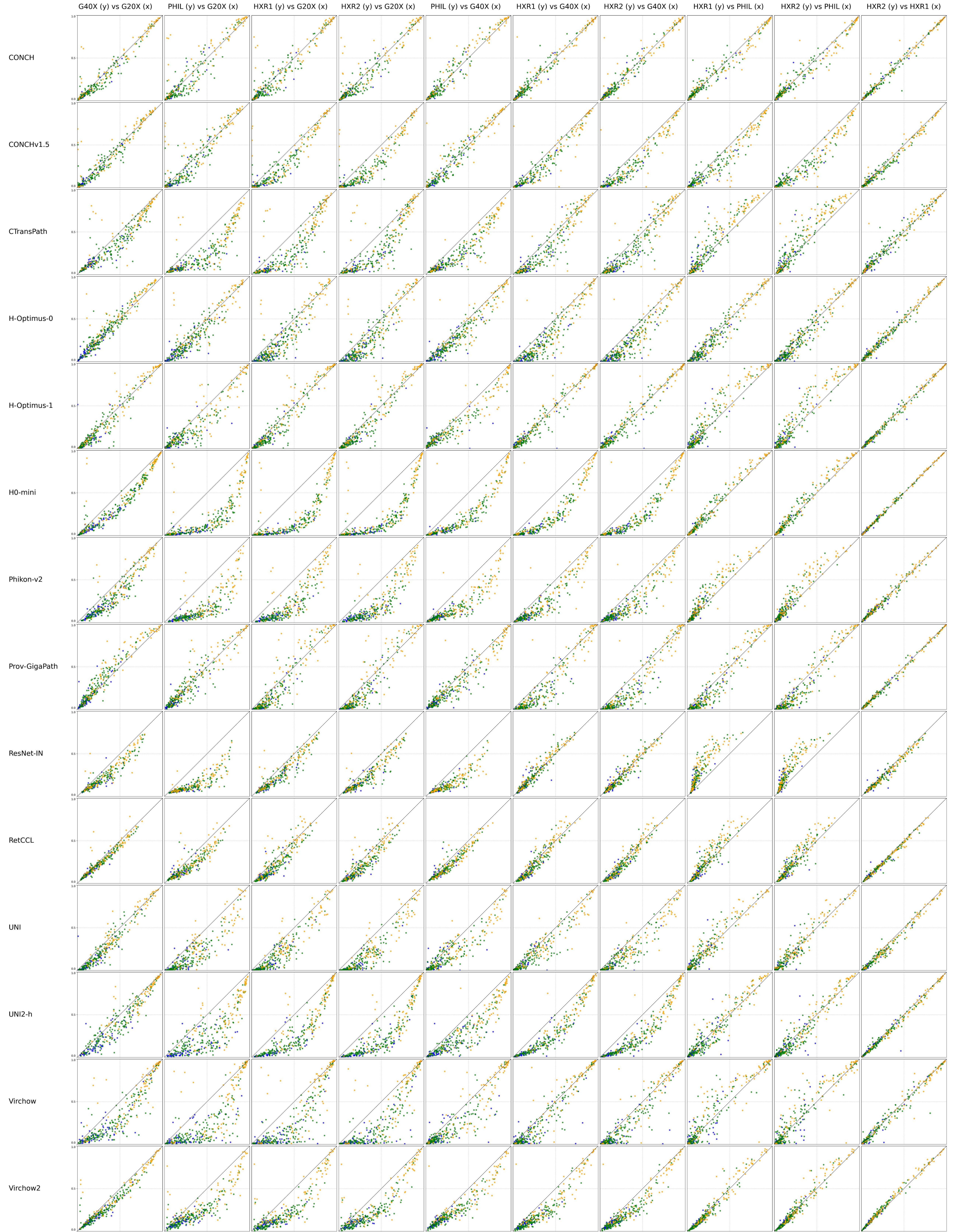}
    \caption{Results for the supervised downstream model evaluation of \emph{calibration stability}, showing examples of scanner-dependent calibration shifts for the multiclass NHG task. This is an extension of Figure~\ref{scatter_conch_h0mini_comparison}, showing results for all evaluated feature extractors (rows).}
    \label{fig:scatter_all_fms_seed7}
\end{figure}

\begin{figure}[t]
    \centering
    \includegraphics[width=0.90\textwidth]{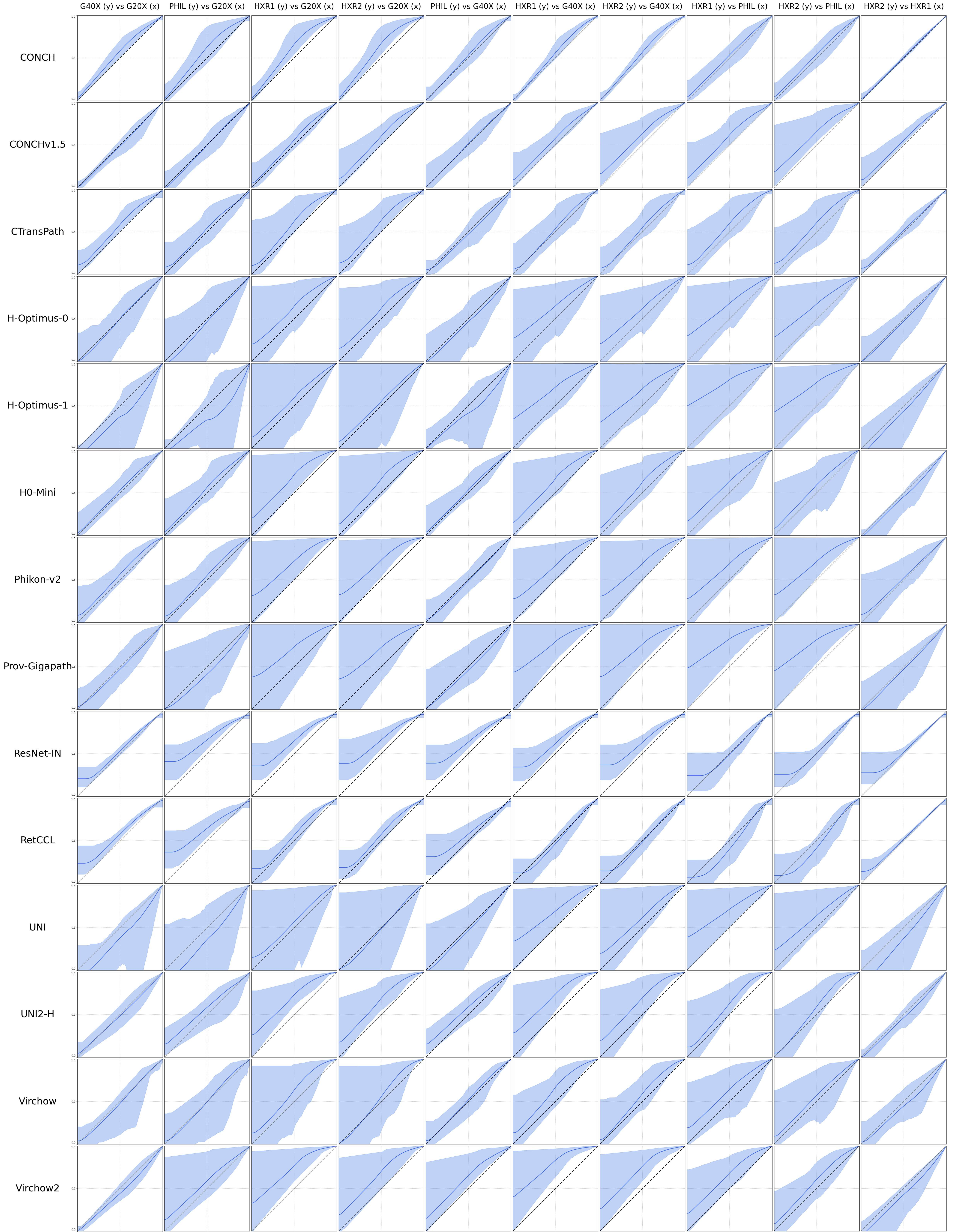}
    \caption{Results for the supervised downstream model evaluation of \emph{calibration stability}, showing LOWESS curves for the \emph{ER status} task. Results are shown for all evaluated feature extractors (rows) across all scanner pairs (columns). This figure follows the same format and interpretation as Figure~\ref{fig:lowess_grade3}, but for the ER status task instead of multiclass NHG.}
    \label{fig:lowess_er}
\end{figure}

\begin{figure}[t]
    \centering
    \includegraphics[width=0.90\textwidth]{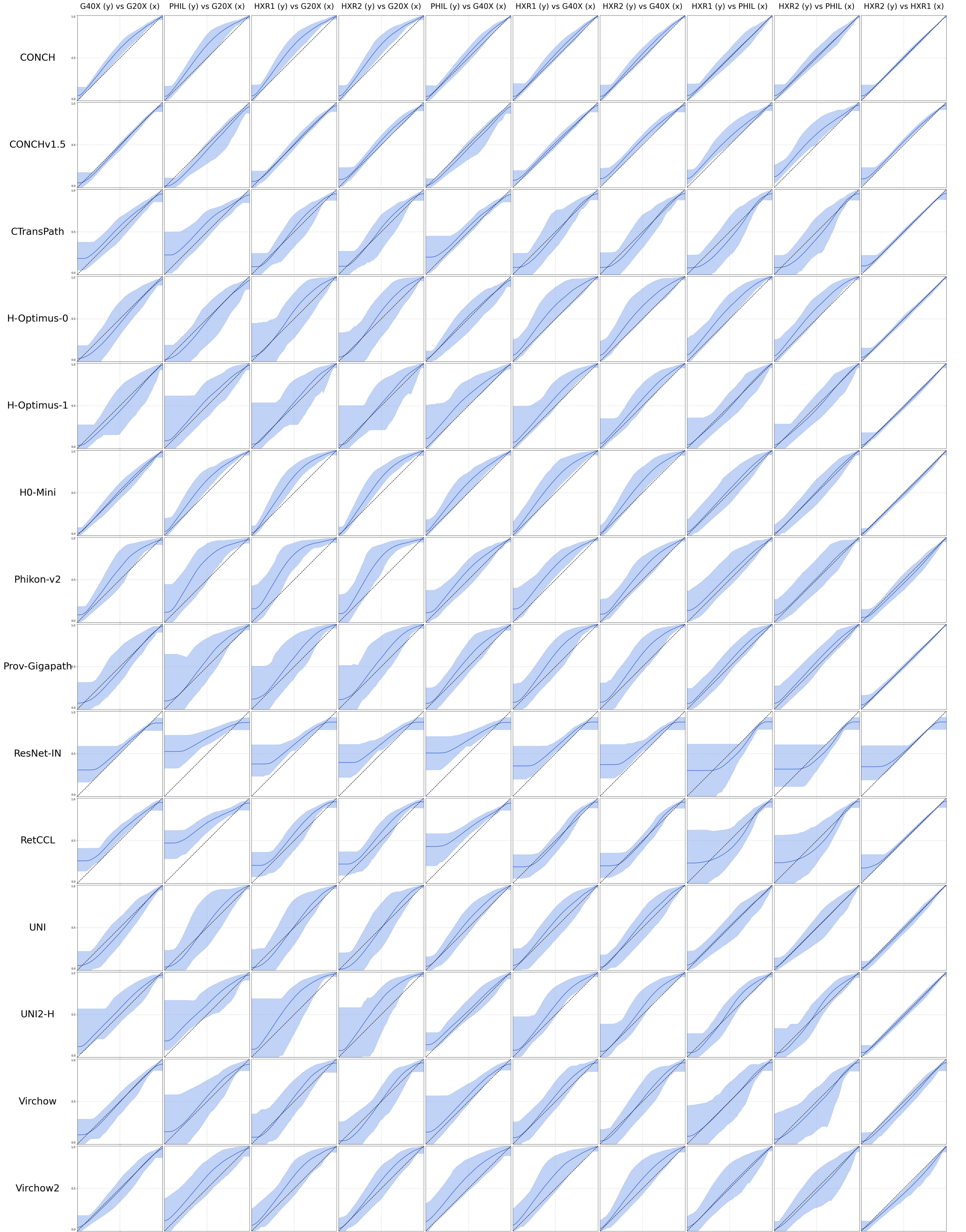}
    \caption{Results for the supervised downstream model evaluation of \emph{calibration stability}, showing LOWESS curves for the \emph{PR status} task. Results are shown for all evaluated feature extractors (rows) across all scanner pairs (columns). This figure follows the same format and interpretation as Figure~\ref{fig:lowess_grade3}, but for the PR status task instead of multiclass NHG.}
    \label{fig:lowess_pr}
\end{figure}

\begin{figure}[t]
    \centering
    \includegraphics[width=0.90\textwidth]{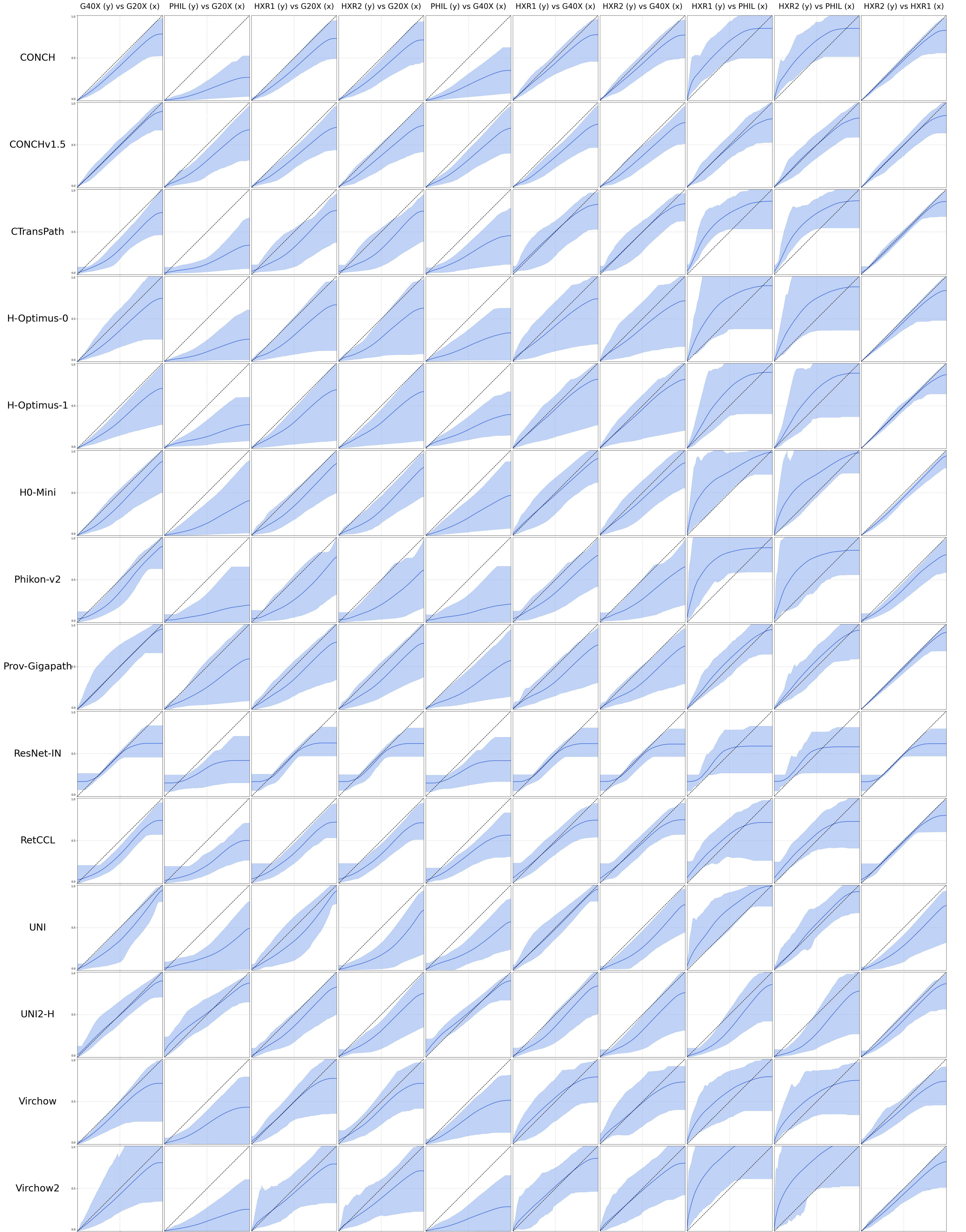}
    \caption{Results for the supervised downstream model evaluation of \emph{calibration stability}, showing LOWESS curves for the \emph{HER2 status} task. Results are shown for all evaluated feature extractors (rows) across all scanner pairs (columns). This figure follows the same format and interpretation as Figure~\ref{fig:lowess_grade3}, but for the HER2 status task instead of multiclass NHG.}
    \label{fig:lowess_her2}
\end{figure}

\begin{figure}[t]
    \centering
    \includegraphics[width=0.90\textwidth]{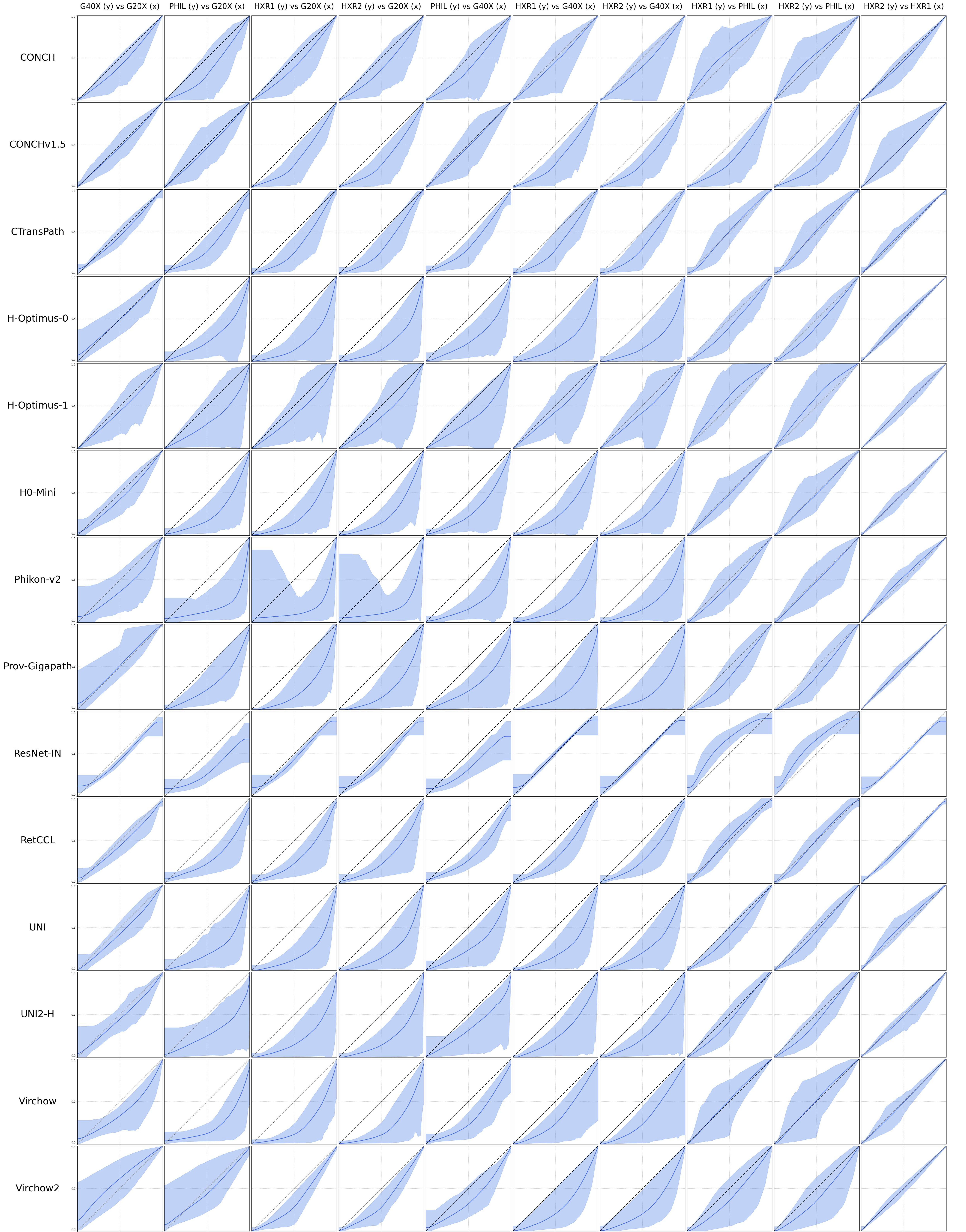}
    \caption{Results for the supervised downstream model evaluation of \emph{calibration stability}, showing LOWESS curves for the \emph{binary NHG (1 vs 3)} task. Results are shown for all evaluated feature extractors (rows) across all scanner pairs (columns). This figure follows the same format and interpretation as Figure~\ref{fig:lowess_grade3}, but for the binary NHG (1 vs 3) task instead of multiclass NHG.}
    \label{fig:lowess_grade2}
\end{figure}

\clearpage
\section{Experimental Details}
\label{appendix:experimental_details}

\begin{table}[ht]
    \centering
    \caption{Hyperparameters used for training ABMIL models on all five clinical tasks in the supervised downstream model evaluation.}\vspace{-2.0mm}
        \begin{tabular}{ll}
        \toprule
        Dropout & 0.25 \\
        Optimizer & AdamW \\
        Initial learning rate & 1e-4 \\
        Weight decay &  1e-5 \\
        Early stopping & After 10 epochs \\
        Bag loss type & Cross-entropy \\
        Maximum epochs & 20 \\
        \bottomrule
        \end{tabular}
    \label{tab:hyperparameters}
\end{table}

\end{document}